\documentclass[sigconf]{acmart}

\usepackage{tikz}
\usepackage{enumitem}
\usepackage{listings}

\usepackage{microtype}
\usepackage{graphicx}
\usepackage{pifont}

\usepackage{xparse}
\usepackage{calc}

\usepackage{colortbl}
\usepackage{xcolor}
\usepackage{array}
\usepackage{multirow}
\usepackage{float}
\usepackage{hyperref}

\definecolor{codegreen}{rgb}{0,0.6,0}
\definecolor{codegray}{rgb}{0.5,0.5,0.5}
\definecolor{codepurple}{rgb}{0.58,0,0.82}
\definecolor{backcolour}{rgb}{0.95,0.95,0.92}

\lstdefinestyle{mystyle}{
    backgroundcolor=\color{backcolour},   
    commentstyle=\color{codegreen},
    keywordstyle=\color{magenta},
    numberstyle=\tiny\color{codegray},
    stringstyle=\color{codepurple},
    basicstyle=\ttfamily\footnotesize,
    breakatwhitespace=false,         
    breaklines=true,                 
    captionpos=b,                    
    keepspaces=true,                 
    numbers=left,                    
    numbersep=5pt,                  
    showspaces=false,                
    showstringspaces=false,
    showtabs=false,                  
    tabsize=2
}

\AtBeginDocument{%
  \providecommand\BibTeX{{%
    \normalfont B\kern-0.5em{\scshape i\kern-0.25em b}\kern-0.8em\TeX}}}

\copyrightyear{2024}
\acmYear{2024}
\setcopyright{acmlicensed}\acmConference[UIST '24]{The 37th Annual ACM Symposium on User Interface Software and Technology}{October 13--16, 2024}{Pittsburgh, PA, USA}
\acmBooktitle{The 37th Annual ACM Symposium on User Interface Software and Technology (UIST '24), October 13--16, 2024, Pittsburgh, PA, USA}
\acmDOI{10.1145/3654777.3676348}
\acmISBN{979-8-4007-0628-8/24/10}

\newcommand{\system}{SHAPE-IT}
\newcommand{\project}{\textit{SHAPE-IT}}
\newcommand{\todo}[1]{}
\newcommand{\qk}[1]{\textcolor{purple}{}} 

\newcommand{\ryo}[1]{}

\newcommand{\customsize}{\fontsize{8}{10}\selectfont}

\definecolor{primitive}{RGB}{0,0,0}
\definecolor{animation}{RGB}{64,150,251}
\definecolor{interaction}{RGB}{255,153,0}

\newcommand{\primitivebox}[1]{%
  \fcolorbox{primitive}{white}{#1}%
}
\newcommand{\animationbox}[1]{%
  \colorbox{animation}{\textcolor{black}{#1}}%
}
\newcommand{\interactionbox}[1]{%
  \colorbox{interaction}{\textcolor{black}{#1}}%
}
\newcommand{\largecircbullet}{%
  \scalebox{1.5}{\textbullet}%
}

\newcommand{\added}[1]{#1}
\newcommand{\removed}[1]{}

\begin{document}

\title{\removed{SHAPE-IT: Exploring Opportunities for Using Text to Author Shape, Motion and Interaction on Shape Displays with LLMs} \added{SHAPE-IT: Exploring Text-to-Shape-Display for Generative Shape-Changing Behaviors with LLMs}}









\author{Wanli Qian}
\authornote{Both authors contributed equally to this research.}
\affiliation{%
  \institution{University of Chicago}
  \city{Chicago}
  \country{USA}
}
\email{michaelq@uchicago.edu}

\author{Chenfeng Gao}
\authornotemark[1]
\affiliation{%
  \institution{University of Chicago}
  \city{Chicago}
  \country{USA}
}
\email{jessegao7@uchicago.edu}

\author{Anup Sathya}
\affiliation{%
  \institution{University of Chicago}
  \city{Chicago}
  \country{USA}
}
\email{anups@uchicago.edu}

\author{Ryo Suzuki}
\affiliation{%
  \institution{University of Colorado Boulder}
  \city{Boulder}
  \country{USA}
}
\affiliation{%
  \institution{University of Calgary}
  \city{Calgary}
  \country{Canada}
}
\email{ryo.suzuki@colorado.edu}

\author{Ken Nakagaki}
\affiliation{%
  \institution{University of Chicago}
  \city{Chicago}
  \country{USA}
}
\email{knakagaki@uchicago.edu}

\renewcommand{\shortauthors}{Qian and Gao, et al.}

\begin{abstract}
This paper introduces \textbf{text-to-shape-display}, a novel approach to generating dynamic shape changes in pin-based shape displays through natural language commands. By leveraging large language models (LLMs) and AI-chaining, our approach allows users to author shape-changing behaviors on demand through text prompts without programming. We describe the foundational aspects necessary for such a system, including the identification of key generative elements (primitive, animation, and interaction) and design requirements to enhance user interaction, based on formative exploration and iterative design processes. Based on these insights, we develop SHAPE-IT, an LLM-based authoring tool for a 24 x 24 shape display, which translates the user's textual command into executable code and allows for quick exploration through a web-based control interface. We evaluate the effectiveness of SHAPE-IT in two ways: 1) performance evaluation and 2) user evaluation (N= 10). The study conclusions highlight the ability to facilitate rapid ideation of a wide range of shape-changing behaviors with AI. However, the findings also expose accuracy-related challenges and limitations, prompting further exploration into refining the framework for leveraging AI to better suit the unique requirements of shape-changing systems.

\end{abstract}

\ccsdesc[500]{Human-centered computing~Interaction devices}

\keywords{Shape Display, Text-based Authoring, LLMs, Code-Generation}

\begin{teaserfigure}
  \includegraphics[width=\textwidth]{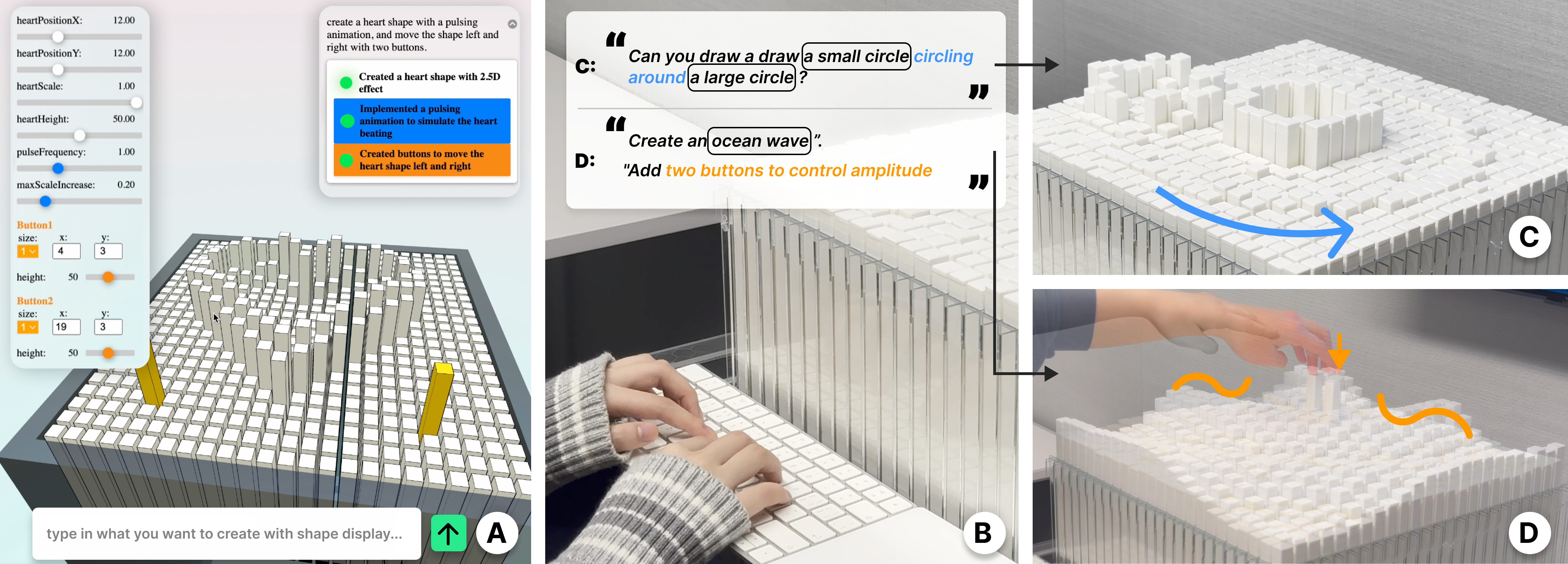}
  \caption{SHAPE-IT lets users to input text to author shape displays. A: SHAPE-IT's GUI. B: a user typing prompts. C: example outcome from our study participant for movement, D: another example for interaction.}
  \Description{Teaser of SHAPE-IT.}
  \label{fig:teaser}
\end{teaserfigure}


\maketitle

\section{Introduction}
\textit{What if our physical environment could dynamically transform through natural language commands?} Imagine a shape-changing table that brings a pen or generates a bookstand by simply asking for it or a floor that transforms itself into a chair or furniture just by saying so. While such ideas have been partially illustrated in research visions~\cite{ishii2012radical, goldstein2004claytronics} or demonstrations~\cite{vink2015transform} to create shape changes based on user intent, most of these dynamic shape configurations currently rely on \textit{pre-programmed} behaviors. How can we create these dynamic shape changes \textit{on-demand} without programming? 

As a first step towards this vision, this paper explores \textbf{text-to-shape-display}, the concept of generating dynamic shapes for pin-based shape displays through natural language commands by leveraging large language models (LLMs). The underlying idea is to employ LLMs to generate executable code that controls each pin of the shape display, allowing users to quickly generate a dynamic shape by simply asking for it. Although the underlying concept is simple, several important questions still remain, such as \largecircbullet{}what common elements text-to-shape-display systems need to generate, \largecircbullet{}what limitations exist when applying basic LLM-based code generation to program shape displays, and \largecircbullet{}how to design such a system from a user-centered perspective. To address these questions, this paper makes three key contributions: 1) identification of \textit{three key generative elements} that emerged from formative exploration, 2) \textit{system requirements and design guidelines} identified via an iterative design process, and 3) \textit{a system design and evaluation} of a proof-of-concept prototype.

First, we identify \primitivebox{Primitive}, \animationbox{Animation}, and \interactionbox{Interaction} as the key elements for the design of text-to-shape-displays. To identify this, we conducted a formative exploration in which we collected 49 representative shape display video clips and recruited 100 crowdsourced participants to investigate how end-users might command the generation of various shape display outputs. Our findings reveal that typical shape display outputs described by users consist of three main components: 1) \primitivebox{Primitive} which defines base geometry, 2) \animationbox{Animation} which describes the movement of primitive objects, and 3) \interactionbox{Interaction} which specifies how user inputs influence primitive objects. Given this result, we propose that text-to-shape-display systems should focus on generating these elements to support a broad spectrum of shape-changing behaviors.

Second, we identified several design requirements for text-to-shape display, including support for iterative refinement through step-by-step authoring, transparency regarding non-deterministic LLM results, and support for real-time exploration via parameter manipulation. These guidelines were established through the development and evaluation of an early prototype, in which we developed a simple LLM-based shape display authoring system, and then conducted a pilot study with ten participants. 

Based on these insights, we developed SHAPE-IT, the first text-to-shape-display system that can control a 24 x 24 pin-based shape display based on text instructions. \removed{Our system consists of several LLM modules}\added{Our system leverages AI-chaining, which connects multiple AI models or tasks in sequence to perform complex operations by leveraging the strengths of each component.}. First, the \textit{prompt-helper} module interprets user input and translates it into three key components: \primitivebox{Primitive}, \animationbox{Animation}, and \interactionbox{Interaction}. Second, the \textit{code-generating modules} take the result of these three decomposed inputs and then generate executable code for each element. \removed{This architecture improves the reliability of code generation through explicit interpretation of user intents.}\added{This AI-chaining architecture improves the reliability of code generation through explicit interpretation of user intents.} In addition, our frontend interface with a conversational UI and shape display simulator allows users to quickly review and modify the shape display behaviors through parameter manipulation or iterative command execution. 

We evaluate our system through two methods: 1) performance evaluation and 2) user evaluation. First, we measure the performance of our system, which results in 82\% compilation success rate, based on 50 random samples gathered from our crowdsourcing study. Second, we invited 10 participants to attend a user study, which included a series of design tasks using our system. We collected qualitative data with semi-structured interviews. Participants' feedback validates the usability and capability of our system. Their creative shape-changing interaction design indicates that SHAPE-IT could facilitate rapid ideation of tangible interaction. We also uncovered the limitations of our system, learned the need for the enhancement of human-AI collaboration in tangible authoring, and the generation of interpretable results. Based on these insights, we discuss how future research could address these issues to broaden this concept to broader shape-changing interfaces.

Finally, our contributions include:
\begin{itemize}
\item \removed{Foundational knowledge,} Requirements, and guidelines for designing text-to-shape displays through formative exploration and iterative development. 
\item \project{}\footnote{Our project code: \href{https://github.com/AxLab-UofC/SHAPE-IT}{https://github.com/AxLab-UofC/SHAPE-IT}}, \removed{the first LLM-based authoring tool for shape displays, utilizing specialized LLM modules to enable the authoring of dynamic shapes, motions, and interactions.}\added{the first LLM-based authoring tool for shape displays uses AI-Chaining, which connects multiple AI models or tasks in sequence to perform complex operations by leveraging the strengths of each component, enabling the authoring of dynamic shapes, motions, and interactions.}
\item Evaluation results and insights highlighting the benefits and challenges of integrating LLMs with shape displays.
\end{itemize}

\section{Related Work}

\subsection{Authoring and control methods for pin-based Shape Display}

To allow lay users to flexibly and intuitively configure shapes, motion, and interaction on pin-based shape displays, researchers have explored various software and hardware implementations aimed at empowering users to design and configure tangible shapes and motions on pin-based shape displays. 
These approaches include pre-planning GUI ~\cite{je2021elevate, yasu2022magneshape, dai2024morphmatrix}, gestural interactions~\cite{everitt2016shapecanvas, hardy2015shapeclip, everitt2016shapecanvas, blackshaw2011recompose, leithinger2014physical}, and tangible instruction based on touch or `scanning' physical objects ~\cite{nakagaki2017animastage, leithinger2014physical, Nakagaki2019inforce}. \project{} enhances the process of designing for pin-based shape displays by enabling users to input instructions in natural language, either by typing or speaking. Unlike traditional methods that are confined to \textit{predetermined options}, our system uses Large Language Models (LLMs) to convert these instructions into code, thereby controlling the display's behavior. This innovation significantly broadens the scope of possible behaviors, moving beyond fixed options and allowing for limitless creativity in shaping display outcomes.

\subsection{Text Authoring and Generative AI in HCI}

Research in using text input to control and author computing systems is a fundamental research realm in HCI, and it is now part of our everyday computing interface, for example, voice assistants~\cite{porcheron2018voice}. Natural language input is one of the most accessible methods for user interaction, and it has great potential for users to achieve complex task goals without needing to remember specific UI elements or buttons~\cite{laput2013pixeltone}, or even programming~\cite{barnett1968natural}.
Text input has been employed in various applications, often combined with other modalities, such as image editing~\cite{laput2013pixeltone}  general point-and-speech UI interaction~\cite{bolt1980put}, and data visualization~\cite{Srinivasan2021}. The accessibility and expressiveness of natural language make it a powerful tool for users to communicate their intentions and achieve desired outcomes in interactive systems~\cite{oviatt2017handbook}, as well as to create and modify interactive Mixed Reality experiences in real-time~\cite{de2024llmr}.

The recent mind-blowing advancements in generative AI using text inputs to generate content have opened up new possibilities for user interaction systems. These tools, which include text-based (ChaptGPT \cite{openai_chatgpt}, Claude\cite{claude3anthropic}), image-based (DALL-E \cite{openai_dalle}), and even video-based models (Sora \cite{videoworldsimulators2024}) are becoming publically available and used world-wide. 

 Following this trend, in the domain of HCI, LLM-based AI generation systems actively being investigated to design novel user interaction harnessing the capability of translating natural language descriptions into generative outputs, including data visualization authoring~\cite{Srinivasan2021}, interactive diagrams~\cite{jiang2023graphologue}, UI prototyping~\cite{Lee2022}, writing assistance~\cite{zhang2023visar}, and end-user programming~\cite{Thayer2022}, creative coding~\cite{angert2023spellburst}, and code generation~\cite{Kazemitabaar2023, ExpectationVsExperience}. To our knowledge, though employing LLM to generate 3D digital models has been explored in the graphics community\cite{michel2021text2mesh, mohammad2022clip,nichol2022point}, translating text into outputs for physical shape-changing devices has not been explored. We find great research opportunities and potential to bridge the realms of human-generative-AI interaction and shape-changing interfaces to create a new thread of research that turns natural language into tangible, dynamic, and responsive shapes. 


Our work advances LLM-based code generation by facilitating dynamic script creation for shape-altering displays and addressing challenges in Generative AI systems. We tackle the discrepancy between user text instructions and the formulation of effective prompts for code generation. As highlighted by ``Why Johnny Can't Prompt'' \cite{zamfirescu2023johnny}, not all users can craft prompts that elicit the desired response from AI systems. Further, Khalid, et al. \cite{Mishra2023PromptingWP} suggest that structured prompts, akin to pseudo-code, enhance code generation accuracy—a principle applicable to text-to-shape display systems requiring precise interpretation of user intent. Despite the growing popularity of AI in code generation, achieving flawless outcomes remains elusive \cite{feng2023investigating}, underscoring the need to minimize errors in such systems. Drawing inspiration from the concept of LLM-Chaining and its potential to improve complex systems \cite{wu2022ai}, we introduce an architecture that preprocesses user commands via LLM-chaining. This approach converts user inputs into structured code instructions based on user intent, thereby enhancing the accuracy of subsequent code generation.

\section{Formative Exploration}
This section describes our formative exploration aimed at informing the design of \textbf{text-to-shape-display} systems. To understand how users construct commands for creating shape-changing behaviors, we conducted a crowdsourced elicitation study. First, we compiled representative video clips of shape displays (N = 49) from previous research papers (N = 21). For each video collected, we asked crowdsourced participants (N = 100) to provide a prompt that could generate the observed motion. By analyzing these user-generated commands (N = 314), we identified common language patterns and key elements of shape-changing behaviors, which will guide the design of our system.

\begin{figure*}[h]
  \centering
  \includegraphics[width=\textwidth]{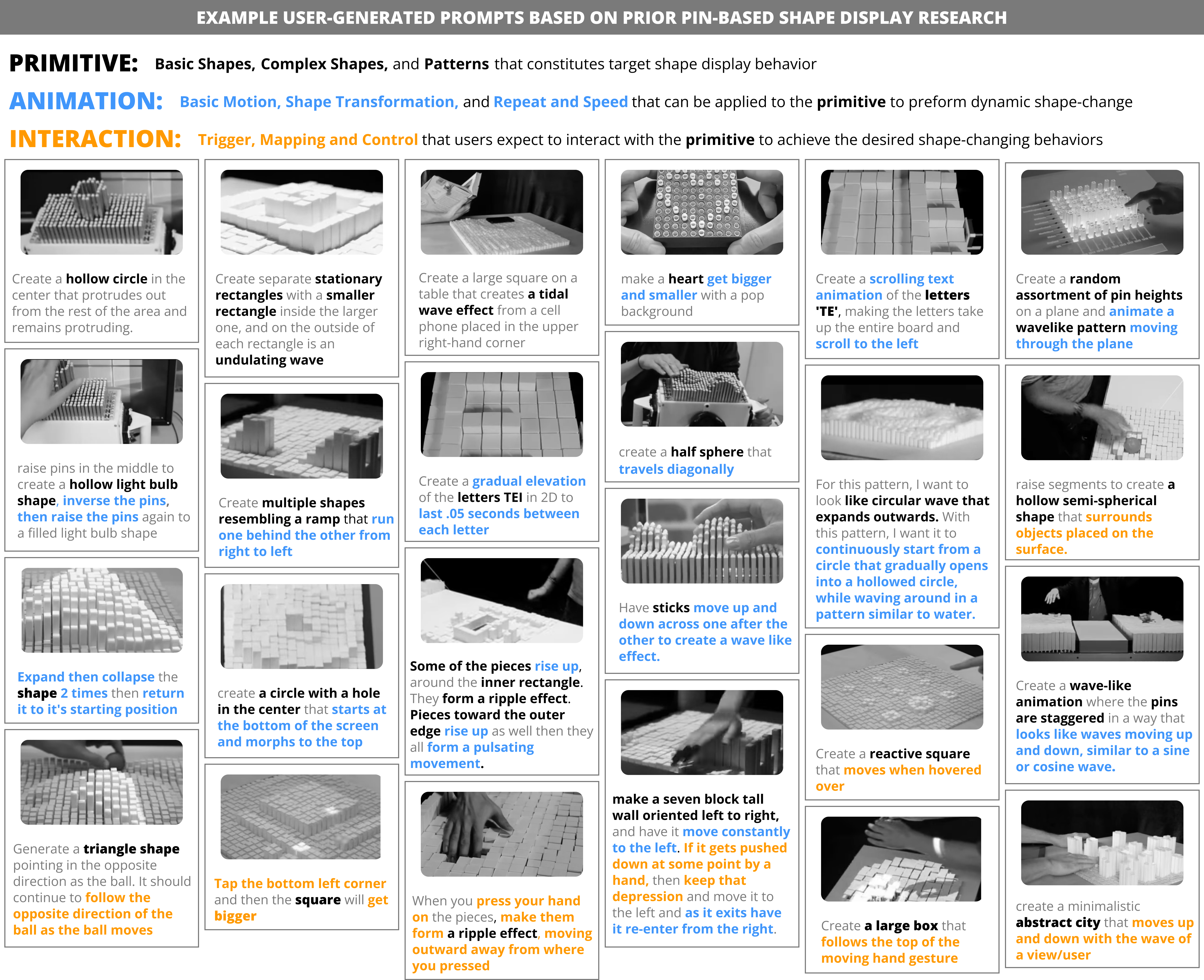}
  \caption{\removed{Primitive, Animation, Interaction elements used in user-generated prompts based on prior pin-based shape display research. Primitive means the fundamental shape or patterns on the shape display that constitutes target shape-changing behavior; Animation stands for the addable behaviors to make primitive perform shape-changing motions; Interaction represents the Primitive that can be triggered or mapped by user interaction control to achieve certain shape-changing behaviors.}\added{Primitive, Animation, and Interaction: Fundamental shapes, motion behaviors, and user-controlled triggers for text-to-shape-display approach learned from prior pin-based shape display research.}}  
  \label{fig:crowdsourcing}
  \Description{formative study results based on prior pin-based shape display research.}
\end{figure*} 

\subsection{Method}
\subsubsection*{\textbf{Dataset}}
For the elicitation study, we collected 49 video clips that represent various behaviors of shape displays. The collection process was conducted in two steps: 1) searching for and collecting relevant research papers, and 2) extracting diverse interactions from video demonstrations. First, we gathered existing research on shape displays by conducting a manual keyword search for \textit{``pin-based shape display''} in the ACM Digital Library, from which we curated those that primarily focused on tangible information displays or shape-changing interactions, excluding works centered on themes such as haptics and actuation techniques. After the selection process (see Appendix~\ref{app:grounding}), we compiled a set of 21 papers ~\cite{vink2015transform, je2021elevate, Siu2019ShapeCAD, suzuki2018dynablock, yasu2022magneshape, leithinger2014physical, Nakagaki2019inforce, follmer2013inform, everitt2016shapecanvas, robinson2016emergeables, nakagaki2016materiable, nakagaki2017animastage, taher2015exploring, leithinger2010relief, leithinger2013sublimate, colter2016soundforms, Siu2018ShapeShift, hardy2015shapeclip, jang2016haptic, strasnick2016applications, ishii2015transform, fitzgerald2018mediate} for our video materials and we extracted short video clips from these research videos. After assembling this pool of clips, we applied several criteria to further filter the data (see Appendix~\ref{app:formative}). These criteria were designed to ensure the essential aspects of the shape-changing behaviors are captured and minimize the presence of external information that could potentially bias or influence participants' decisions. This process resulted in a final set of 49 representative video clips\footnote{A link to the playlist can be found:  \href{https://www.youtube.com/playlist?list=PLjuhl7c_2gWjDHpYBVObHveGpM7NpMUOj}{study materials}}, each with an average length of 4.8 seconds.

\subsubsection*{\textbf{Participants}}
Participants for this study were recruited through Prolific, an online platform that provides academic researchers with access to a diverse pool of participants. During recruitment, we did not apply strict demographic criteria or target specific user groups, with the exception of requiring English as the primary language. This requirement ensured that participants could provide detailed text inputs in English, which was essential for our study. We recruited a total of 100 participants, compensating them at an average rate of \$12 per hour, which amounted to an average of \$1.60 per participant. For detailed procedures, see Appendix~\ref{app:formative}.

\subsubsection*{\textbf{Task}}
We asked each participant to generate a prompt for each of four video clips. The task was divided into two parts: First, during the introduction and training phase, participants were presented with a video (a heart shape scaling up and moving from the top right to the middle in the TRANSFORM~\cite{vink2015transform} video) and two example prompts to ensure they understood that the task was to provide instructions rather than mere descriptions. Second, in the main task, participants viewed four randomly selected video clips from the curated 49 videos. They were then asked to type their own instructions for each clip. From 100 participants, a total of 314 valid\footnote{prompts like ``I don't know how to describe'' or it doesn't focus on shape-changing effect are treated as invalid.} user instructions were collected from 400 raw data points.

\subsubsection*{\textbf{Coding Analysis}}
Given the 314 user-generated prompts, two of the authors conducted a thorough thematic analysis~\cite{braun2006using} to identify commonly used language patterns and recurring themes. The analysis began with an initial coding phase, where each author independently examined a subset of the prompts and assigned preliminary codes to capture the salient features and concepts expressed by the users. After several rounds of analysis, refinement, and discussion within the research team, we reached a consensus on the final categorization scheme, which consisted of three main generative elements: \textbf{\textit{primitive}}, \textbf{\textit{animation}}, and \textbf{\textit{interaction}}. Figure~\ref{fig:crowdsourcing} illustrates a set of example videos, associated user-generated prompts, and color-coding of these key elements. In the following sections, we describe each element in detail. These identified elements inform the design of our system, detailed in Section 5.

\subsection{Primitive}
We observed that participants often referenced a specific 'object' on the shape display to describe shape-changing effects. We define \primitivebox{Primitive} as the basic geometry of that constitutes target shape display behavior, which can be either basic shapes, geometries, or movement patterns. These primitives exclude added animations or interactions, considering them accessory features. Primitives may include motion, such as a ``wave'' pattern. Categories of primitives identified are \textbf{\textit{Basic Shapes}} (geometric shapes like circles, squares, and triangles), \textbf{\textit{Complex Shapes}} (more detailed forms like letters, icons, and landscapes), and \textbf{\textit{Patterns or Basic Layouts}} (specified arrangements or effects, often described when precise shapes cannot be articulated). These serve as foundational elements for constructing more intricate forms and behaviors.

\subsection{Animation}
\animationbox{Animation} elements are defined as enhancements to \textit{primitives}, altering geometry parameters to achieve dynamic and continuous motion. These elements introduce various motion behaviors to primitives, including \textbf{\textit{basic motion}} (e.g., translations like rising, falling, and pulsating movements), \textbf{\textit{transformations}} (shape or state changes such as morphing and revealing), and \textbf{\textit{speed and timing}} (adjustments from fast to slow). Additionally, \textbf{\textit{Repetition and Speed}} are crucial, with users frequently requesting repetitive or looping animations, dictating the rhythm and periodicity of the motion. These animation features, applied atop primitives, facilitate dynamic shape transformations.

\subsection{Interaction}
The \interactionbox{Interaction} element focuses on how users expect to interact with and control the primitive shape or pattern to achieve the desired shape-changing behaviors. Such as \textbf{\textit{trigger}}: Users specify various triggers that initiate or influence the shape-changing effects, including physical touch, pressing on the shape displays, hovering above the shape displays; \textbf{\textit{mapping and control}}: Users describe different mapping mechanisms for continuously interacting with the shapes. This includes direct and indirect manipulation and parameterization. These interaction control techniques allow users to manipulate and customize the shape-changing effects by being applied to primitives.

\section{Early Prototype and Pilot Study}
Our goal is to identify the challenges and potential benefits of an LLM-based \textbf{text-to-shape-display} system. To understand the user experience of such a system, we developed an initial working prototype to conduct a pilot study, based on Buchenau's experience prototyping protocol~\cite{buchenau2000experience}. This helps us gain insights from the user's firsthand experiences by allowing them to directly engage with functional systems. 

\subsection{\textbf{Early Prototype}}
Our initial prototype consisted of two main components: 1) a frontend shape display simulator built with Unity, and 2) a backend LLM-based code generator built with Python. The backend code generator leverages the GPT-4 API to translate users' textual instructions into executable Unity code. To ensure the generation of executable code, we employ few-shot prompting \cite{promptingguide2023fewshot} techniques\footnote{Few-shot prompting in the context of Large Language Models (LLMs) involves providing the model with a small set of examples (typically one to five) to guide it on a specific task. These examples, comprising input-output pairs, act as a mini-dataset for the model to adapt its responses accordingly. }. For example, each prompt provides a collection of example commands, such as \textit{``create a heart shape''} or \textit{``Created a ball shape and a button to control toggling appear and disappear of the ball shape''}, along with their corresponding author-prepared Unity programs that are designed to control the height of each pin in a 30 x 30 shape display simulator (see Appendix~\ref{app:early-prompt} for more prompt details). The frontend interface features simple and basic UI elements, including a text input box, an area for displaying AI responses, and a shape display simulator. After receiving a response from GPT-4 API, the backend system sends the output to the frontend interface with the JSON format, so that the system can run the GPT-generated program to animate the 30 x 30 shape display simulator with a single input text box.

\begin{figure}[h]
  \centering
 \includegraphics[width=1\linewidth]{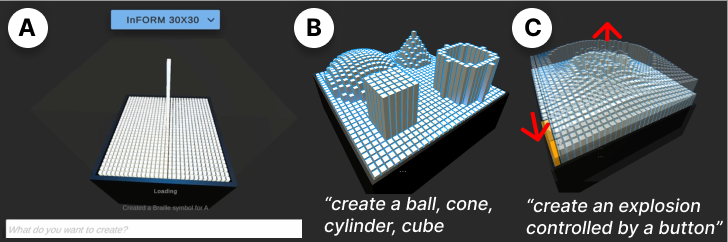}
  \caption{Early UI prototype (A) for creating basic geometries (B), and interaction (C).}  
  \label{fig:early_prototype}
  \Description{early Unity-based prototype, and two examples of generated results}
\end{figure}

\subsection{Pilot Study}
We conducted a pilot study with 10 participants (6 males, 4 females), including 4 experts in programming, 5 in design, and 1 in product management, with their ages ranging from 22 to 33 (average age 26). First, participants were given a brief introduction and a walkthrough of the system. They were then tasked with performing a free-form ideation activity (40 minutes) to develop their desired concepts using our system, followed by a short interview (20 minutes) to discuss the system's usability and gather suggestions. Throughout the study, we observed the participants' interactions and collected feedback. Each session lasted approximately 60 minutes, and each participant was compensated \$15 for their participation.

\subsection{Findings in Challenges and Informing Feature Improvments}
Overall, participants recognized the potential of the text-to-shape-display approach, as all of them could construct certain shape-display behaviors using texts easily. However, we also found multiple key problems that have to be resolved to build \textbf{text-to-shape-display} system, that is easy to use, employ and iterate repeatedly. We elaborate on them in five key insights below, which we incorporated for SHAPE-IT implementation, detailed in section 5. 


 \largecircbullet{}\textbf{Enabling Micro-Adjustments:} The only option for users to create or modify the shape display behavior was prompting with the pilot frontend UI. Even very simple minor adjustments (e.g., tuning the height of a rectangle or controlling the wave motion speed) required re-prompting, which required typing a follow-up prompt and waiting approximately 40-50 seconds for GPT agent processing, only to get another result which may still not satisfy user's intent. To address this issue of \textit{difficulty of quick micro-adjustments}, we propose to make \textbf{parameter generation, and adjustments via GUI} as one of the required features for the SHAPE-IT system. In such a system the LLM agents can generate a set of parameters that can be tuned and adjusted by the users to for quick modifications.

\largecircbullet{}\textbf{Accessing Conversation History:} Participants also expressed frustration with the \textit{inability to review and compare and revert to previously generated results}, especially when the newly created ones were worse than before. This motivated us to implement a feature of \textbf{conversation history} that allows users not only to view the prior prompts and generated codes, but also to revert to any of the history results, and re-prompt based on that prior result.

\largecircbullet{}\textbf{Transparency of AI generation: } During the study, participants expressed concerns that the process between user input and AI output appeared as a ``black box,'' leading to feelings of distrust and frustration, particularly when the system failed to meet their expectations. This feedback highlights the essential need to develop an AI system that \textbf{transparently communicates} its reasoning process, provides feedback, and reports errors to users. Such improvements are crucial for fostering a more effective human-AI interaction by enhancing transparency throughout the generation process.

\largecircbullet{}\textbf{Token Economy\footnote{
In the context of Large Language Models (LLMs), we define ``token economy'' as efficient management and utilization of the input and output tokens within the constraints of the model's processing capacity. Tokens, in this context, are the basic units of text that the model processes, which can be words, parts of words, or punctuation marks, depending on the tokenization process used by the model.} and Context management:} the pilot system uses a naive context management strategy that compounds user inputs and code responses in LLM conversations in each user interaction to enable context-based editing which leads to a quadratic increase in input token costs during the conversation. Given each iteration the context is compounded with a code result which has massive content, this indicates poor token economy and ineffcient context preservation strategy. This issue is worsened by feeding multiple few-shot prompting examples to the system, further inflating input length and model context capacity after 3-5 interaction rounds for participants during the study. Hence, in SHAPE-IT framework, we utilize \textbf{Retrieval-Augmented Generation (RAG)} \footnote{RAG enhances language models by integrating information retrieval(fetching relevant documents in response to queries and using these to guide output.), allowing for more accurate and information-rich text generation.} and \textbf{Multi-Agent framework} to provide a better token economy while maintaining necessary contexts for code generation ensuring consistent user experience. 

\largecircbullet{}\textbf{The Need for Component Segmentation:} Our pilot study highlighted that users often request additional features without intending to modify existing elements. For example, requests like ``make it move left and right'' or ``create a button to control the height'' are common. However, our initial system, lacking component segmentation, generated new scripts that not only addressed these requests but also included comprehensive definitions extending to the user's entire context. This usually resulted in unintended changes to the primitive elements, and high variability exists in code results. This issue underscores the need to separate primitives, animations, and interactions within our system. Our formative study findings align with this insight that animations and interaction enhancements should be built on and added to existing primitives, facilitating incremental development based on an existing base.

In Section 5, we discuss how we apply the proposed features learned from the Pilot Study in detail.

\section{SHAPE-IT}
We developed SHAPE-IT, a system that turns user instructions into executable code for creating and modifying shapes. Similar to the initial prototype, our system has two components: 1) a frontend interface for the shape display simulator built with JavaScript and 2) a backend LLM-based code generator built with Python. Our system is designed to address all of the five design requirements identified in the early prototype. In this section, we describe the system design and implementation of \system{}.

\begin{figure*}[h]
  \centering
  \includegraphics[width=1\linewidth]{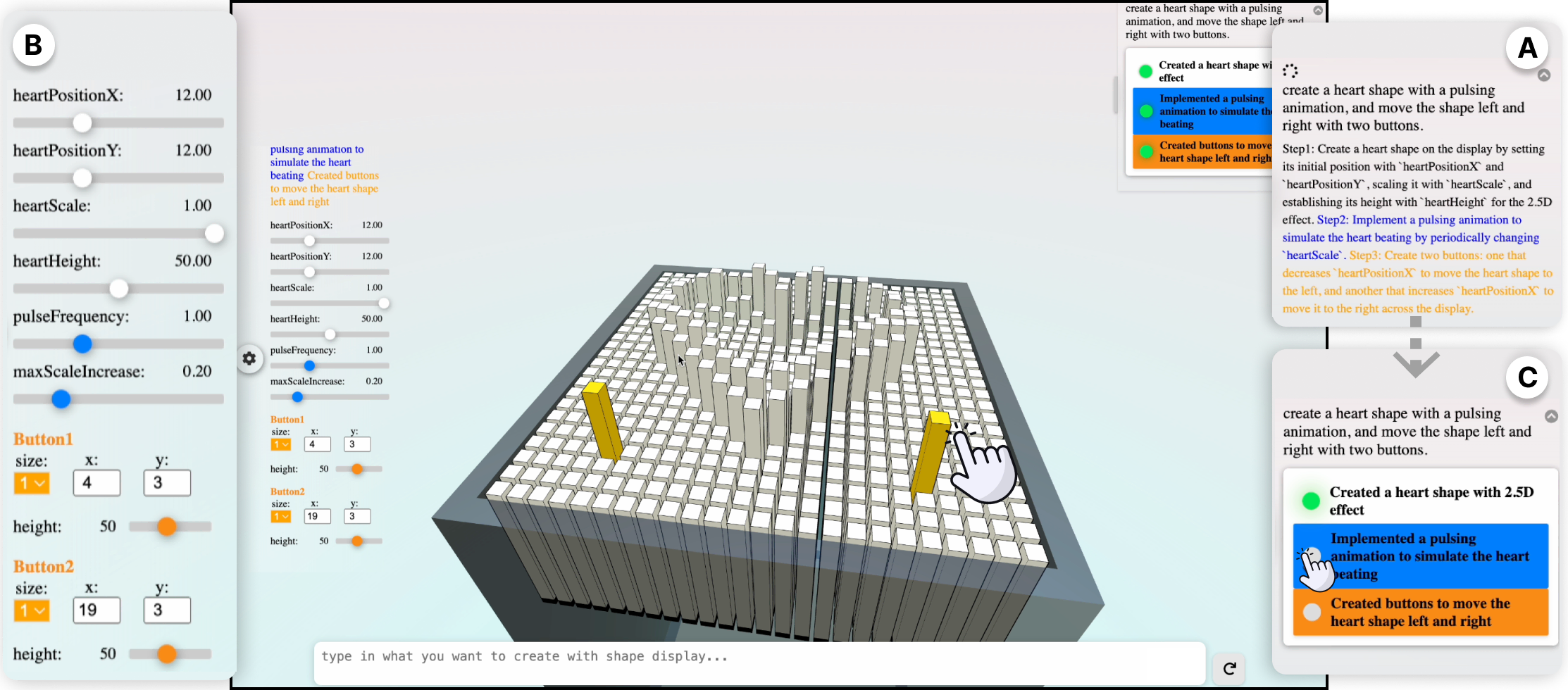}
  \caption{SHAPE-IT UI with an example prompt; (A) The feedback message communicates the generating steps to users, indicating \primitivebox{Primitive}, an \animationbox{Animation}, and an \interactionbox{Interaction} scripts are generated. (B) the Parameter Control Panel; (C) Loaded scripts appear as Script Card UIs, where they can be toggled on/off.}  

  \label{fig:example-walkthrough}
  \Description{SHAPE-IT UI}
\end{figure*}

\begin{figure*}[h]
  \centering
  \includegraphics[width=1\textwidth]{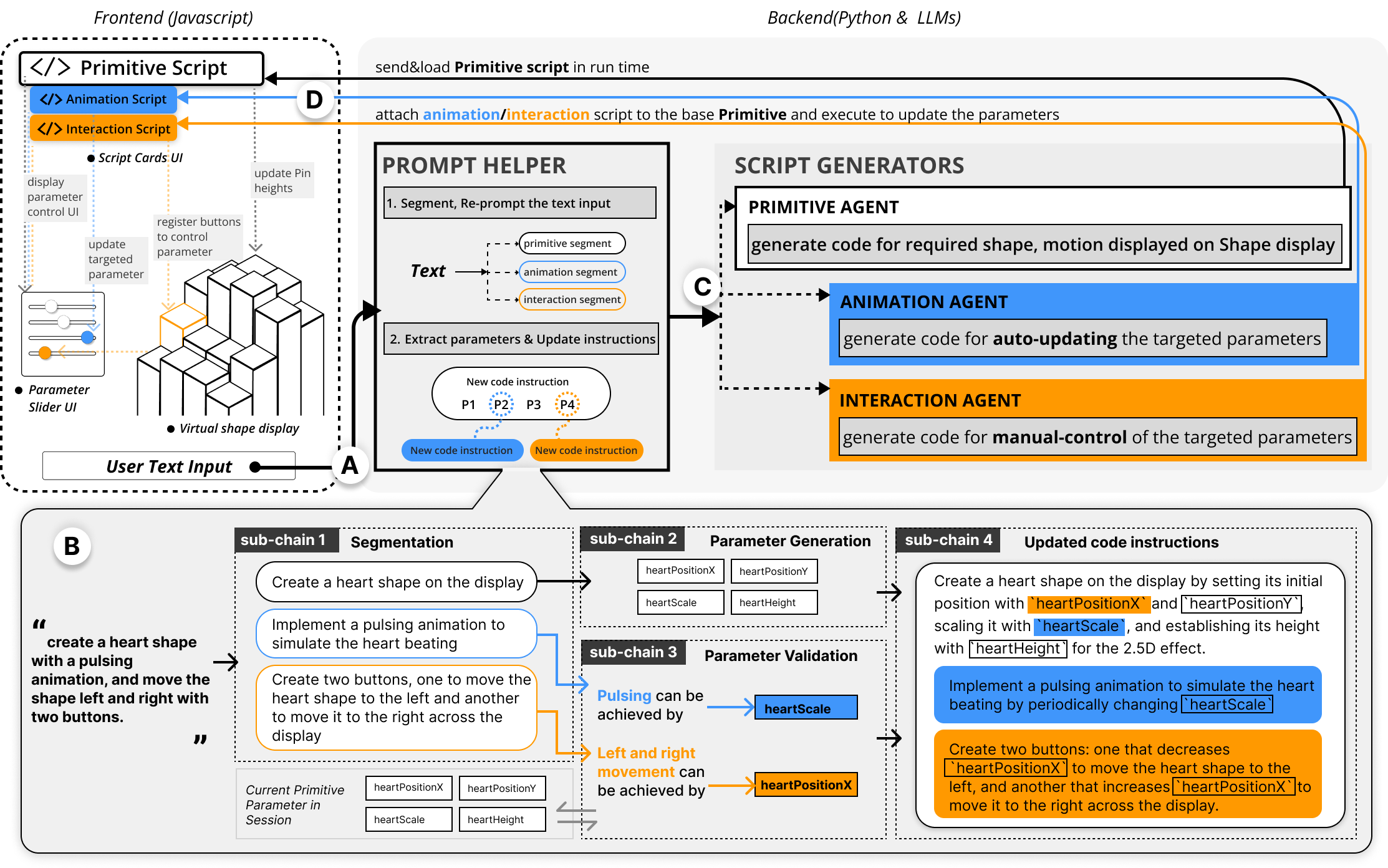}
  \caption{SHAPE-IT System Architecture. (A) Text Input is handed to Prompt Helper, (B) Prompt Helper processes it to provide code instructions; (C) Script Generators generate the Javascript codes based on the instructions; (D) the generated codes are executed in the frontend, controlling the shape display and creating UI elements for users to adjust parameters.}  
  \label{fig:system-architecture}
  \Description{System Architecture}
\end{figure*}

\subsection{System Walkthrough}

SHAPE-IT not only allows users to provide instructions in an incremental manner, where users are able to do follow-up commands to add extra behaviors or make edits on current creation, but also can understand a composite prompt to create dynamic and interactive shape-changing behaviors at once. we describe how our system works through the following example prompt: \textit{``create a heart shape with a pulsing animation, and move the shape left and right with two buttons.''}. Our system generates the interactive animation as seen in Figure~\ref{fig:example-walkthrough}. Below, we detail how our LLM pipeline produces the resulting outcome based on this user input.

\subsubsection*{\textbf{Step 1. Segmentation: Decompose User Input into Three Key Elements}}
First, the system decomposes the user input into three key elements: \primitivebox{Primitive}, \animationbox{Animation}, and \interactionbox{Interaction}, informed by our formative exploration (Figure~\ref{fig:system-architecture} A, B1). In this example, the system segments the inputs as follows: 1) Primitive: \primitivebox{\textit{``Create a heart shape on the display''}}, 2) Animation: \animationbox{\textit{``Implement a pulsing animation to simulate the heart beating''}}, and 3) Interaction: \interactionbox{\textit{``Create two buttons, one to move the heart shape to t-}} \interactionbox{\textit{he left and another to move it to the right across the display''}}.

\subsubsection*{\textbf{Step 2. Parameter Generation: Identify Key Shape Properties for Dynamic Motion}}
In our system, dynamic motion is achieved by updating the shape properties of primitive objects. To this end, the system identifies the necessary parameters for the specified animation and interaction  (Figure~\ref{fig:system-architecture} B2-3). For instance, in the current example, the system identifies four key parameters: 1) \textit{positionX} of the heart shape, required for horizontal movement, 2) \textit{positionY} of the heart shape, required for vertical movement,  2) \textit{scale} of the heart shape, which is necessary for the pulsing animation, and 4) \textit{height} of the heart shape to control the visibility. The system uses these properties as controllable parameters for both the animation and interaction. 

\subsubsection*{\textbf{Step 3. Parameter Validation: Ensure integrity of parameter control on the auxiliary components}}
SHAPE-IT ensures the integrity of parameter control within the animation and interaction segments by incorporating an additional parameter validation step. This step specifically assesses whether the current primitive parameters can fulfill the objectives outlined in the animation and interaction segments. This is crucial for managing scenarios where a user's subsequent actions may necessitate an animation or interaction that implies the need for a parameter not included in the initial primitive setup. However, in this context, the code generation process in Step 2 has already taken into account the requirements of all three segments. Consequently, it has generated a list of parameters that encompasses all necessary implications, leading to successful validation.

\subsubsection*{\textbf{Step 4. Update Code Instruction: Prepare Proper Code Instructions by Incorporating Extracted Parameters}}

Next, the system constructs three code instructions by incorporating extracted parameters (Figure~\ref{fig:system-architecture} B4). For instance, in the current example, the resulting prompts are as follows, where \textit{[params]} are identified parameters: 1) Primitive: \textit{``Create a heart shape on the display by setting its initial position with [heartPositionX] and [heartPositionY], scaling it with [heartScale], and establishing its height with [heartHeight] for the 2.5D effect.''}. 2) Animation: \textit{``Implement a pulsing animation to simulate the heart beating by periodically changing [heartScale]''}. 3) Interaction: \textit{``Create two buttons: one that decreases [heartPositionX] to move the heart shape to the left, and another that increases [heartPositionX] to move it to the right across the display.''}. These instructions will be sent to the frontend as well, notifying the users of the creating tasks(shown in Figure~\ref{fig:example-walkthrough} A).

\subsubsection*{\textbf{Step 5. Script Generation: Generate Executable Code for Primitive, Animation, and Interaction}} 
Given the updated prompt, the system generates the script for each component. The system includes three independent LLM modules, each responsible for constructing a script for primitive, animation, and interaction (Figure~\ref{fig:system-architecture} C). These LLM modules generate executable code that controls the height of each pin in the shape display. The prompt engineering for this module combines three techniques: 1) rule-based prompting, 2) retrieval augmented generation (RAG), and 3) structuring of input and output, which will be described in a later section.

\subsubsection*{\textbf{Step 6. Code Execution: Run the Generated Script for Shape Construction and Animation}}
Once the backend system generates the executable script, it sends the information to the frontend in JSON format, which generates a dynamic shape in a Three.js simulator which will be synced to a physical 24 x 24 shape display (Figure~\ref{fig:system-architecture} D). The frontend system first processes the primitive script to construct static primitive shapes. Then, it executes the animation and interaction scripts to make the dynamic motion. The extracted parameters are also displayed as a list of slider values in the user interface. When the animation or interaction alters these parameters, the slider values are updated accordingly. Simultaneously, the frontend interface shows the previous prompt, generated code, and the system's explanations of how the provided prompt was interpreted.

\subsubsection*{\textbf{Step 7. Interaction and Modification: Interact with the Shape Display and Update Results}}
Once the code is executed, the user can see the result, interact with the generated shape, or refine the outcomes by adjusting the generated parameters. The slider values are not only useful for the animation and interaction modules but also enable users to interactively modify the shapes without having to issue new commands. Users can also update the result by entering a new command. For example, if a user says, \textit{``instead of moving the position, I want it to rotate when I click the button''}, the system adapts by adding \textit{[heartRotation]} parameters in the next round, where the system modifies the code for the existing primitive and animation objects accordingly.

\subsection{Implementation}
\subsubsection*{\textbf{Backend System}}
The backend of the system consists of two components: a \textit{prompt helper} module (Steps 1-3, Figure~\ref{fig:system-architecture} A-B) and a \textit{script generator} module (Step 4, Figure~\ref{fig:system-architecture} D). We provide our prompts for both prompt helper and script generator modules in Appendix~\ref{app:new-prompt }.

\subsubsection{Prompt Helper}
We developed a prompt helper module to convert natural language inputs into appropriate code instructions. The prompt helper is built upon GPT-4-Turbo and utilizes AI-chaining method~\cite{wu2022ai} that sequences multiple language models to progressively transform user instructions into executable code directives. This approach breaks down the overarching task into smaller, specialized steps. In our implementation, this module functions through four sub-chains: 1) segmentation, 2) parameter generation, 3) parameter validation, and 4) code instruction prompt generation. After generating segments and parameters, the parameter validation chain assesses whether the identified parameters align with the user's intent. If not, it initiates a subroutine to adjust the existing parameters and correct the command based on the revised parameters. These parameters are then used to refine the segmentation results, culminating in the formation of code instructions derived from the generated parameters.

\subsubsection{Script Generator}
Script generator modules take updated prompts as input and generate executable code as output. There are three script generator modules for primitives, animations, and interactions, all built with GPT-3.5-Turbo-0125. We selected this model to achieve faster code generation. The prompt engineering for the script generator is based on the following three techniques: 1) rule-based prompting, 2) retrieval-augmented generation (RAG), and 3) input/output structuring. 

\noindent
\textbf{\textit{(1) Rule-Based Prompting:}}
Rule-based prompting establishes clear guidelines for the generated results. It emphasizes functional code that aligns with our specific requirements by providing the following four instructions: structured code templating through the use of parameter lists for initialization, coding practices to ensure modularity and system compatibility, detailed instructions for error handling, and a focus on goal orientation to enhance code complexity and efficiency.

\noindent
\textbf{\textit{(2) Retrieval Augmented Generation:}}
We employ a retrieval-augmented generation (RAG)\cite{python_langchain_retrieval} to match instruction-code pairs relevant to the given instruction. We have developed three distinct input-output collections categorized under primitives, animations, and interactions. These collections are stored within separate chroma vector stores for efficient retrieval. During the retrieval phase, we use the \textit{semantic similarity example selector}\cite{langchain_semantic_similarity_selector} from the Langchain framework\cite{langchain_homepage} to identify the three most semantically related instructions in our database. These are then incorporated into the final generation template. This method ensures that the generated code leverages relevant examples and adheres to the formatting and structural guidelines established by our rule-based prompting. \added{Each input-output collection for \primitivebox{Primitive}, \animationbox{Animation}, and \interactionbox{Interaction}(detailed in Appendix~\ref{subsec:primitive_collection}, \ref{subsec:animation_collection}, \ref{subsec:interaction_collection}) was built under our formative study results (section 3.2). This formative study categorization guided the development of multiple code examples for each category. Additionally, we included new examples based on observations of frequently attempted shape primitives by participants in our pilot study. }


\noindent
\textbf{\textit{(3) Input/Output Structuring:}}
For JSON output, we utilize the \textit{structured output parser}\cite{langchain_python_output_parsers} within the same LangChain framework. The output JSON follows this structure: 1) user input, 2) parameters needed for code generation, 3) the category of the script generator, 4) a system-generated explanation for interpretation, and 5) the executable code generated by the script generator.

\subsubsection{Context Management}
SHAPE-IT introduces a method for managing context that addresses the challenges related to the token economy observed in our pilot system and, in the meantime, facilitates continuous conversations between users and the system. Our framework employs code generation agents to hold onto a single history of output code as short-term memory. This allows for consistent and incremental user modifications. Additionally, we incorporate a Retrieval-Augmented Generation (RAG) mechanism within each code generator, enabling the selective retrieval of relevant database examples, which diminishes the necessity for extensive few-shot examples. Furthermore, the Prompt Helper Agent is responsible for managing long-term memory, archiving all user inputs along with their associated code instructions, thereby preserving the comprehensive context of the interaction. Despite the quadratic increase in token cost for the Prompt Helper Agent, it is much more compact than code results. Yet, by doing this, we allow code-generator agents to sustain a linear growth in token cost. Therefore, we offer an improved context management strategy with a better token economy than the pilot system without breaching the model context limitation.

\subsubsection*{\textbf{Frontend Sytem}}
The frontend system consists of 1) a shape display simulator and control interface, 2) conversation history, 3) an error handling console, and 4) hardware communication. 

\subsubsection{Shape Display Simulator and Control Interface}
The frontend interface renders 24 x 24 shape display simulator built with ThreeJS, along with sliders for the generated parameters. The initial values of these parameters are scaled up by a factor of three and reduced by a third to establish the upper and lower bounds of the slider range. Through real-time manipulation of these sliders, users can directly control the primitives. For interactions, interactive elements like buttons identified in the interaction script are highlighted in orange. Each button element features a dropdown list for selecting its size (1x1 or 2x2), as well as sliders for adjusting its position (posX and posY) on the display and height.

\subsubsection{Conversation History}
The UI records and displays the entire history of interactions between the user and the system. Each generated script is represented as a card with a textual description, color-coded based on their types. Users can navigate back to any previous card to continue editing from that point, facilitated by the backend's ability to roll back to a specific history state. This feature not only provides a comprehensive overview of the authoring session but also enables users to reference previous commands, inputs, and outcomes.

\subsubsection{Error Handling Console}
When the front end encounters compilation errors while loading the generated scripts, it requests the backend to regenerate the scripts again. \removed{The script-generating module keeps a record of the most recent script output, allowing it to recognize and attempt to circumvent the error upon retrying. This strategy ensures that the system adaptively responds to errors, minimizing disruptions to the user's creative process. }

\added{Specifically, when the front-end encounters a compilation error, it sends a 'compile error message' to the backend LLM modules for each module to regenerate the code. Since each code generator retains the memory of the previously generated result, the modules can adaptively correct the errors. This is a common practice for handling errors in code generation. \cite{chen2023teaching}.}

\subsubsection{Hardware Communication}
For the physical shape display, we replicated a 24x24 (434 × 434mm display area, with 100mm stroke length) inFORM shape display, based on an existing design ~\cite{cooperHewittInform}. For bi-directional data transmission, we utilized the MQTT (Message Queuing Telemetry Transport) protocol. The MQTT server, hosted on our university's server, was built using RabbitMQ. Our frontend JavaScript application can publish and subscribe to MQTT topics, which convey the target heights of pins based on the generated script. It also receives real-time height information of the pins from OpenFrameworks, which is used to control the movements of hardware motors. All these hardware development and communication details are elaborated in prior publications~\cite{follmer2013inform, nakagaki2016materiable}.

\section{Performance Evaluation}
\subsection{Method}
\removed{To assess the performance of our system, we conducted an evaluation using a subset of data from our crowdsourcing study in formative exploration. Specifically, we randomly selected 50 data points from a total of 314 user-generated prompts(shown in Appendix~\ref{app:Evaluation_prompts}). These prompts were then input into our system to determine its success rates, defined by the system's ability to attach generated code to the scene without resulting in compile errors.}

\added{To assess the performance of our system, we conducted an evaluation using a subset of data from our crowdsourcing study in formative exploration. Specifically, we randomly selected 50 data points from a total of 314 user-generated prompts (shown in Appendix~\ref{app:Evaluation_prompts}). These prompts were then input into our system to determine its success rates, defined by the system's ability to attach generated code to the scene without resulting in compile errors. The system architecture listed for comparison marks the progression from our initial system from the pilot study to the current SHAPE-IT system. Our hypothesis is that the addition of modules transitioning from the baseline to the SHAPE-IT system should not negatively influence the code compilation success rate. Maintaining or increasing the compilation success rate validates this hypothesis while also achieving the goal of a more intuitive user-generation flow for SHAPE-IT. This evaluation method was inspired by LLMR \cite{de2024llmr}.
}

The success rate was calculated using the following formula:
\[
\bar{S} = \frac{1}{n} \sum_{i=1}^{n} \left( \frac{1}{m_i} \sum_{j=1}^{m_i} s_{ij} \right), n = 50
\]
In this formula, $s_{ij}$ represents the success or failure of each segment within a sample as a binary outcome (1 for success, 0 for failure). The average success for each segment yields the sample's success rate. We then computed the overall average success rate by taking the mean of these rates across all samples.

Our evaluation spanned four models, two of which is a baseline single code-generator architecture without segmentation based on (primitive, animation and interaction) intent segmentation, and the latter two is Multi code-generator architecture that involves intent segmentation.\footnote{Baseline and Baseline+RAG models, are single code-generator architecture without segmentation, ${m_i} = 1$. In contrast, the SHAPE-IT Segmentation and SHAPE-IT segments user prompts into three distinct segments—primitive, animation, and interaction—thus ${m_i} = 3$ for these systems.}:
\begin{itemize}
    \item \textbf{The Baseline model}, which utilizes a Single code-generator with few-shot prompting. Where we feed the user input directly to the code generator with few-shot selected examples from the combination of Primitive, Animation, and Interaction Agents.
    \item \textbf{Baseline+RAG}, Single code-generator baseline, replacing few-shot prompting with Retriever-Augmented Generation (RAG). with RAG retrieval collection built with combined example of Primitive Animation and Interaction Examples 
    \item \textbf{SHAPE-IT Segmentation}, Multi-code generator framework which does simple segmentation without parameter generation and inference. Similar architecture with current system Figure~\ref{fig:system-architecture} with lesser version of PromptHelper (Chain 1 only). 
    \item The \textbf{SHAPE-IT} System, our current system Figure~\ref{fig:system-architecture} which builds upon the Segmentation model by adding a parameter pipeline and code instruction.
\end{itemize}

Additionally, we evaluated model latency to inform our choice of code-generating agent, using the baseline system as a reference.



\begin{figure}[h]
  \centering
  \includegraphics[width=1\linewidth]{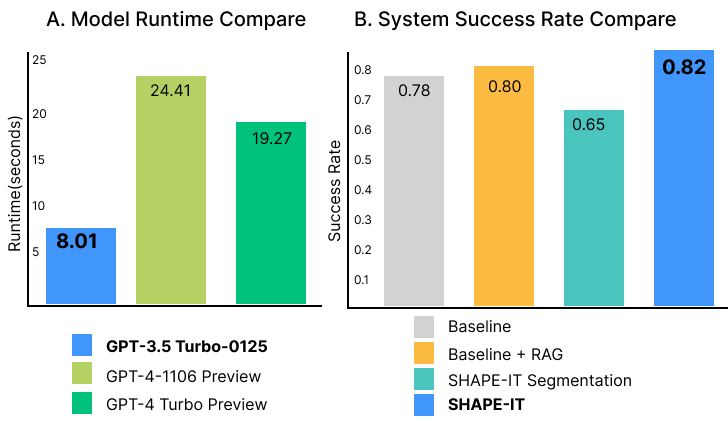}
  \caption{A.Model Runtime Comparison. B.System Success Rate Comparison}  
  \label{fig:model_comparison}
  \Description{Tech evaluation bar chart}
\end{figure}

In evaluating the latency of code generation, we compared three widely used large language models (LLMs) tailored for code-generation tasks: gpt-3.5-turbo-0125, gpt-4-1106-preview, and gpt-4-turbo-preview. Our findings, illustrated in Figure \ref{fig:model_comparison} A., revealed that gpt-3.5-turbo-0125 exhibited significantly lower latency, averaging 8.01 seconds across 50 samples. This was compared to 24.41 seconds for gpt-4-1106-preview and 19.27 seconds for gpt-4-turbo-preview. Consequently, we selected gpt-3.5-turbo-0125 as the base model for our code-generation agents due to its efficiency.

From Figure\ref{fig:model_comparison} B., the Baseline system achieved a success rate of 78\%, and Baseline+RAG achieved 80\%. Suggesting moderate compilation success for our baseline model. The SHAPE-IT Segmentation system has a lower success rate of 65\%, due to the lack of coherece in construction of segments. While the segmentation results reasonably reflect user's intent, we observe that the failure cases mainly comes from the animation/interaction segments doesn't provide a coherent instruction with the primitive. SHAPE-IT system addresses this problem by introducing parameter inference after the segmentation process to allow code-instruction to be built for the three segments leveraging the inferred parameters, this approach allows each segments to be interweaved by parameter instructions therefore provides coherence structuring of the segments It achieves the highest success rate of 82\%, suggesting that the compilation of the code is no longer effected by this incoherence.

\section{User Evaluation}

We conducted a study involving 10 participants who have never interacted with shape displays to assess the effectiveness of SHAPE-IT. Our aim was to identify both its strengths and limitations, informing potential areas for improvement through example interaction processes and results.


\subsection{Participants}

Our participants, comprising 6 males and 4 females with an average age of 28.89, were non-experts in shape displays but had familiarity with LLMs. Coding proficiency varied, with 7 out of 10 participants reporting proficiency, while 4 were skilled in design and 8 were familiar with prompting strategies. Each participant spent 1 hour in the study and received \$15 for their time.

\begin{figure*}[h]
  \centering
  \includegraphics[width=1\linewidth]{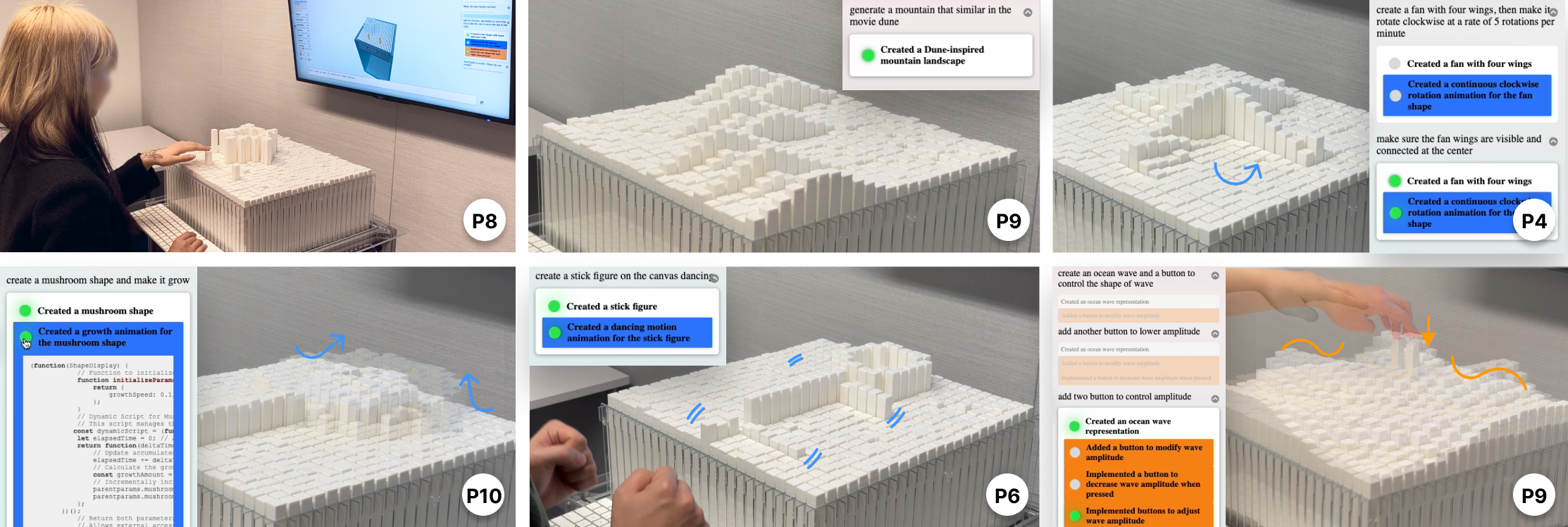}
  \caption{Study setup and examples of participants' creations: 
  P8 created an interaction to use a button to move the position of a star shape.
  P9 generated an animated mountain terrain from a movie. 
  P4 created a spinning fan. 
  P10 animated a growing mushroom. 
  P6 made a dancing stick figure.
  P10 designed a controllable wave animation with buttons.}  
  \label{fig:user-examples}
  \Description{Participants results from the study}
\end{figure*}

\subsection{Procedures and Tasks}

Participants underwent an introductory session (10 mins) where one of the authors explained the study procedures followed by three main tasks (35 mins).

\noindent\textbf{Task 1:} Participants were tasked with creating a simple primitive shape and then enhancing it by incorporating additional animation or interaction behaviors. This task tests the system's ability to enable user input with the intention to author animation/interaction based on an existing primitive.

\noindent\textbf{Task 2:} Participants were asked to provide composite instructions, combining multiple commands into a single sentence. This task aimed to test the system's ability to parse and execute complex instructions.

\noindent\textbf{Task 3:} Participants engaged in rapid idea exploration, where they were free to utilize any instructions to create the desired shape-changing behaviors. This task allowed for creative exploration of the system's capabilities.

\noindent Finally, we conducted a semi-structured interview (15 mins) to gather insights from the participants' perspectives.



\subsection{Results}
Using the video recordings and the transcripts from the interviews, we synthesize a few qualitative insights. Some of these insights are specifically applicable to our system and others represent ongoing challenges with generative AI as a whole.

\subsubsection{\added{Overall Generated Results}}
\added{
Based on participants' instructions, SHAPE-IT successfully created basic shapes (e.g., circles, cubes, triangles) and some simple animations (e.g., Star Twinkle, Raining effect). Impressively, it can create a flower with 8 petals, where the size of the petals can be controlled by sliders. It also achieved a chasing animation, where participants could use the UI to control the position of the chased object, while another object actively chased it. However, more complex shapes (e.g., Four-Wheel Car, Cat Face) and animations (e.g., Morphing, Eye Blinking) often failed. Interestingly, the system worked for prompts like ``circling around'', but failed with ``Orbiting''. Simple interactions (e.g., button-triggered text) were generally successful, while intricate interactions (e.g., controlling speed) frequently failed.
}
\added{
Additionally, we have included a table in the Appendix \ref{app:summary_of_success_user_study} summarizing all attempted prompts by participants, with a subjective assessment of the results (Success, Partial Success, Failure). This provides a glanceable overview of the user study outcomes.
}


\subsubsection{Overall impressions}

The study indicated strong interest in the system, with notable praise for the user interface. Participants commended the system's adeptness at interpreting language inputs and offering granular control over parameters. They observed that the system's segmentation of primitives, animations, and interactions corresponded closely with their expectations, also suggesting that the system's initial feedback output accurately reflected their intentions. However, this raises expectations for the system's subsequent visual outputs(via code results) towards which participants commonly expressed a desire for greater accuracy.

Participants highly valued the ability to adjust parameters alongside 3D content and found the physical, interactive buttons on the shape display to be a novel experience. However, in some instances, participants found the parameters to be overly complex -- such as when attempting to create a human face -- where the adjustable parameters were overwhelmingly extensive. This provokes questions about potentially limiting the number of adjustable parameters in the interface to reduce cognitive overload.

\subsubsection{Concrete vs Abstract Creations}

During Task 3, participants encountered a tension between concrete shapes (squares, cubes, letters, emojis, and mathematical surfaces) and more abstract shapes (zombies, cats, and birds). When abstract shapes were desired, they often found the quality of the generated shapes to be unsatisfactory. This tension is an expected challenge inherent in generative AI. For example, while imagining a cube typically yields a consistent mental image among a broad population, abstract shapes like zombies or cats vary widely in interpretation. In our case, this challenge is further compounded by the resolution limitations of the shape display itself.  
\added{We recommend that future researchers develop code examples for various primitive shape categories. This approach will help GPT achieve better code generation by following predefined code structures, thereby improving the quality of the desired shapes.}

\subsubsection{Apparent Learning Curve}

Participants noted that the conceptual models of primitives, animations, and interactions didn't always seamlessly align with their use cases. For instance, the inability to spawn a primitive through interaction is a functionality our system does not support because it considers animations and interactions as secondary to a single primitive. They expressed a recurring desire for more transparency about the system's structure during prompting, acknowledging the challenge of understanding these supposed constraints up front, especially for a novice user. While there's a common belief that natural language input requires minimal training, this underscores the presence of a learning curve associated with such interfaces, especially when the system contains additional mechanisms that improve the experience in the long term (such as parameter adjustment).

\subsubsection{Prompting vs. Coding} 

When comparing user preferences between the coding interface and LLM authoring, feedback varied significantly based on individuals' coding proficiency. Participants without coding experience found the LLM tool beneficial for intuitive exploration and design iteration. However, participants with coding skills preferred using both the coding interface and the LLM system together. This hybrid approach allowed for precise adjustments, especially when initial outputs didn't match their intentions. Notably, one proficient coder viewed the LLM tool skeptically, preferring a more hands-on coding approach. This highlights the challenge of accommodating both novice and expert users in authoring tools~\cite{cockburn2014supporting, bailly2023computational, angert2023spellburst}. \added{Future systems could offer options for expert users to directly edit and refine the generated code, as indicated by the study findings.}

\subsubsection{The Importance of AI Interpretability}

Participants value the step-by-step feedback from the AI system, which enhances understanding of its decision-making process. They appreciate how adjustments and feedback influence the resulting 3D content, leading to a sense of achievement when they get the result they intended. Conversely, discrepancies between the AI-generated content and participants' expectations can cause confusion and frustration. Participants prefer upfront recognition of system limitations and transparent communication to set realistic expectations and mitigate disappointment. This highlights the importance of advances in AI interpretability. 

\subsubsection{Future Prospects for AI Authoring: Text-to-Shape Display vs. Other LLM Tools}

During our discussions on the evolution of AI authoring tools, participants expressed optimism about the future of these innovations. A common thread in these conversations was the admiration for the tangible aspects of AI-generated content, an area where many current AI authoring tools fall short. The ability to physically interact with AI-generated 3D objects was highlighted as a novel and valuable feature, setting our text-to-shape display system apart from other LLMs.

Many users recognized the unique strengths of various AI authoring tools, but they particularly appreciated the tangible interaction offered by our system. One user eloquently states, ``Compared to SORA, I can physically feel the 3D creations and even use tangible buttons for control, which is cool. This could be incredibly beneficial for 3D designers.'' This feedback underscores the potential impact and utility of tangibility in AI authoring, suggesting a promising direction for further development in making digital creations more accessible and interactive.

\section{Potential Application Spaces}
\begin{figure}[h]
  \centering
  \includegraphics[width=1\linewidth]{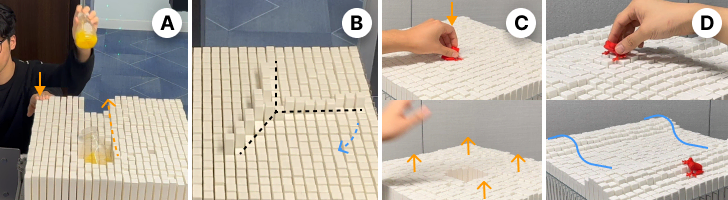}
  \caption{Potential applications that are enabled by SHAPE-IT: A: pass a beverage across the table by a hidden button; B: tangible clock displaying 7:02; C: a button that acts as a trigger to trap a character; D: flush a character by host manipulation of the scene of an emerging wave;}
  \label{fig:applications}
  \Description{Application ideas}
\end{figure}

By allowing users to author dynamic shape-changing behaviors on pin-based shape displays through simple text instructions. Here, we briefly introduce how SHAPE-IT could be employed in applications such as adaptive tabletop furniture and gaming \& storytelling. The results shown in Figure~\ref{fig:applications} were actually generated from our SHAPE-IT system.

\noindent\textbf{Adaptive Tabletop Furniture:}
Inspired by the vision of adaptive and dynamic furniture presented in TRANSFORM~\cite{vink2015transform}, where shape-changing tabletop surfaces dynamically react to users, SHAPE-IT brings this vision close to reality by enabling users to create customizable and interactive tabletop features only by speech. For instance, as illustrated in Figure~\ref{fig:applications} A, a user can create a cup holder with a hidden button that allows for adjusting the holder's position. They can also flexibly ask the display to create a room interior, as B represents a shape-changing clock generated based on language instruction.
  
\noindent\textbf{Gaming and Storytelling: }
SHAPE-IT enables users to provide real-time instruction, which can expand the entertainment and expressive capability of the shape displays. For example, board games like Dungeon \& Dragons require the host to use static game sheets and verbal communication as a way to bring the player an immersive game experience. SHAPE-IT understands such a natural language-based instruction and creates on-demand dynamic scenes on shape displays that align with host descriptions, which enhance their gaming experiences. As illustrated in Figure~\ref{fig:applications}C, the host can author a trap mechanism that is triggered when a player figure steps on a specific trigger, forming a cage that traps the frog character, which could be flexibly generated based on a player's narrative design. The game host can also dynamically alter the game terrain based on the storyline, providing a customizable gaming experience. Figure~\ref{fig:applications}D showcases an example where the frog character sitting on the island is caught off-guard by a rising tide. By not requiring any programming experience, SHAPE-IT allows players to plan and implement game mechanisms or tangible story design via text input. 

\section{Limitation \& Future Work}
In this paper, we explored \textbf{text-to-shape-display} system to allow users to author shape display behavior through natural language instruction, using LLMs. Our proof-of-concept system has shed light on numerous research opportunities, as a first attempt to configure shape-changing interfaces using text-based instruction. This section reflects on our research, discussing the limitations and future work. It also includes system/UI design recommendations for future researchers in both the AI/LLM domain and shape-changing interface domain to explore this novel research area.


\subsection{Prototype Limitation \& Potential Improvements}

\subsubsection{Flexible Authoring Framework}
Our existing system is structured to facilitate the generation of a single primitive, supplemented by a framework that supports animation and interaction enhancements. This design choice stems primarily from the inherent limitations in our code's ability to bind animation and interaction functionalities exclusively to specific primitives. Through user feedback, it became evident that such a setup is restrictive, especially when users wish to generate primitives through interactive or animated processes. To cater to a wider array of design possibilities, we recognize the imperative need to develop a more versatile framework. By allowing animations and interactions to be freely associated with any element, thereby enabling the construction of complex, non-linear scenes, we can enhance the system's utility and flexibility for diverse design applications.

\subsubsection{Parameter Bound Control Issue}
In our prototype, we acknowledged the absence of a robust mechanism for inferring parameter bounds. To achieve a sophisticated inference of parameter bounds, it is essential to conduct a meticulous analysis of system constraints, map out the dependencies between parameters, and utilize data-driven insights alongside machine learning techniques. This approach will facilitate the accurate determination of parameter limits in line with operational constraints and dependencies. Incorporating robust validation methods and a feedback loop is crucial for refining the accuracy of these inferences over time. Furthermore, designing this system with scalability and flexibility in mind will ensure it can adapt to evolving system designs and external factors, significantly bolstering the prototype's functionality and reliability in diverse operational contexts.

\subsubsection{Deploying and Improving the Prototype as Usable Tools}

Our intent for our prototype system is to be an accessible, generalizable tool for HCI researchers, designers, or even novices to easily plan and customize shape display projects as we open-source the code. To this end, there are other improvements we could make to the system to be further generalizable, for example, to adapt to shape displays with different configurations (resolution, number of pins, pin stroke, etc.), as our prototype was confined for the specific 24 x 24 shape display hardware. By making it accessible online with any browser, we consider our tools useful not only for designing behavior on shape display hardware but also for informing people considering/planning to develop shape display to identify their hardware requirements by testing target behaviors interactively via natural languages.

\subsubsection{\added{Physical Constraints of the System}}

\added{The physical constraints of our system are largely divided into two types of geometries:}

\added{
\noindent\textbf{Impossible Geometries:} These include geometries that are not feasible due to hardware limitations, such as floating or overhanging structures.
}

\added{
\noindent\textbf{Intractable Geometries:} These refer to complex geometries that are challenging to handle, such as the animation of an object rotating along the lateral axis (as opposed to the vertical axis).
}

\added{ While the constraints in category (a) are purely due to hardware limitations that the LLM cannot handle, those in category (b) can be addressed by improving the LLM's prompting capabilities beyond simple 'x-y array' interpretations. This includes incorporating a 3D mesh renderer/shader and enhancing the code structure to manage complex geometries more effectively.
}


\begin{figure}[h]
  \centering
 \includegraphics[width=1\linewidth]{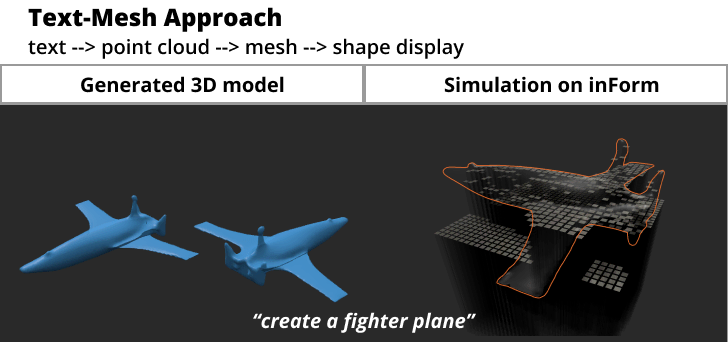 }
  \caption{ Text-to-point-cloud prototype's example result}  
  \label{fig:point-e}
  \Description{early try with test-to-pointcloud approach.}
\end{figure}

\subsection{Potential Future Direction and Implication for the General Approach}

\subsubsection{Limitation of Code Generation, and Potentials of other LLM Tools}

In light of the previously discussed limitation, while the code-generation LLM approach in our system is very good at generating numerically describable motion and geometries (e.g., wave, square, cone), it suffers from making semantic shapes and motion (e.g., car model, human face, etc.) In the computer graphics community, AI-assisted LLM-based geometry generation systems are well-explored ~\cite{mohammad2022clip, Michel_2022, shechter2022neuralmls, li2022ganimator}

In fact, in the early stage of our project, we experimented with employing the text-to-point-cloud LLM tool, Point-E~\cite{nichol2022point}, to generate a 3D model using text to be rendered on the shape displays (Figure \ref{fig:point-e}). While this prototype was capable of generating semantic models, we found the AI system not interpreting the shape display configuration (vertically moving arrays of pins) had an issue when representing some shapes (e.g. granular detail shapes). Also, it is limited in creating motion and interaction. Hence, we have employed a code generation approach for our final approach.


We believe the future \textit{text-to-shape-display} system should integrate multiple types of LLM tools to mitigate each pros and cons. In such a future system, code-generation LLM should have access to generate geometric 3D models that can be employed in the executed code to supplement the weakness of creating semantic shapes. Other LLM approaches, such as generating images or videos, which can be translated into pin height of shape displays, could be another option, as image and video generation tools are actively advancing to flexibly create images in a fast manner \cite{DALLE2}. Still, we consider code-generation LLM should be central for shape-changing displays (or shape-changing interfaces, in general), as the behavior of these hardware is usually confined to computationally controlled code.


\subsubsection{Incoprating Multi-Modal Interactions}

Additional future direction for the system should be in incorporating rich multi-modal interactions, to harness the tangibility and embodied affordance of the shape display. Beyond text input, such a system could incorporate gesture and speech interaction \cite{bolt1980put}, allowing users to point at a location of the display and summon a shape. Tangible interaction could be incorporated to edit (or manipulate) the behavior generated by the system to author them in a physical manner, fusing the interaction technique explored in prior works \cite{leithinger2014physical, follmer2013inform, blackshaw2011recompose}. Such an approach should require an additional AI agent in our system so it can flexibly interpret the gestures and tangible cues. Such a direction would help us reach to the vision of reconfigurable material or clay \cite{ishii2012radical, goldstein2005programmable}, that reconfigure based on user intent expressed in tangible, gestural, and speech interaction.

\section{Conclusion}
In this paper, we introduced SHAPE-IT, a \textbf{text-to-shape-display} system, harnessing multiple LLM agents (GPT) to generate code to control pin-based shape display based on text inputs by users. The system allows users to provide natural language instruction to author shape, motion, and interaction of the hardware. Our early exploration, based on the crowdsourcing study and an early prototype, employing a single LLM agent, informed us how people describe behaviors on shape displays using texts and how our system should be designed, including UI features and multi-agent architecture. Our implemented system was evaluated through a technical evaluation, which reported the code-compilation success rate, and a user study, which revealed the usability of the system, as well as diverse tangible and shape-changing results created from participants' text input. We consider this paper to be the first step in opening up a novel research realm in AI-infused shape-changing UIs that can dynamically shift shapes by flexibly understanding users' intent.

\begin{acks}
\added{We acknowledge the CERES program and Prof. Andrew Chien at the University of Chicago for their support. We thank Richard Liu and Rana Hanocka from 3DL at the University of Chicago for their support for the initial exploration of this research. 

We also appreciate the help of Actuated Experience Lab members with the project, specifically Ran Zhou, for advising the User Study methods. Lastly, we thank everyone who helped with the assembly process of the inForm shape display (originally designed in MIT's Tangible Media Group), including Chenfeng's mom, Lilith Yu, You Li, Yifan Zhou, Willa Yang, Matthew Jeung, Emilie Faracci, Harrison Dong, Raymond Qian, Ramarko Bhattacharya, and Skylar Liu.}
\end{acks}

\bibliographystyle{ACM-Reference-Format}
\bibliography{references}

\appendix
\section{Formative Exploration Materials}
\label{app:grounding}

\subsection{Inclusion Criteria for Prior Pin-Based Shape Display Research Selection}
 This inclusive approach allowed us to capture a broader range of shape-changing behaviors and applications while maintaining a focus on displays capable of rendering rich visual information beyond solely haptic or tactile feedback. To ensure a cohesive and targeted analysis, we defined the following inclusion criteria(C) for the reviewed papers:

\begin{enumerate}[label=\textbf{C.\arabic*},wide=0pt]
    \item \textbf{Device Specifications:}
    \begin{itemize}[noitemsep, topsep=0pt, parsep=0pt, leftmargin=*]
        \item \textbf{Actuation Mechanism:} Devices must utilize a pin-based shape display with linear motion along the z-axis. Exclusions: inflatable matrices (e.g., Flat Panel Haptics), devices with bi-stable movement (e.g., certain braille displays), devices with additional degrees of freedom (e.g., PolySurface, TiltDisplay), and modular shape displays (e.g., ShapeClip, ShapeBots, HapticBots).
        \item \textbf{Device Orientation and Mapping:} The shape display must be oriented on a flat x-y plane. Exclusions: devices like HaptiVec, Real-Time Physical Prototyping Tools based on Shape-Changing Displays, and PoCoPo. Only vertically moving pin-based displays are considered.
        \item \textbf{Size and Resolution:} Devices should not be excessively small (e.g., fingertip or watch size) or overly large (e.g., room-scale). Exclusions: devices simplistically display shapes with few pins (e.g., LiftTiles, TilePop). Inclusions: devices with higher resolution capable of detailed shape representation (e.g., Elevate).
    \end{itemize}
    
    \item \textbf{Research Purpose:}
    \begin{itemize}[noitemsep, topsep=0pt, parsep=0pt, leftmargin=*]
        \item \textbf{Focus of Research:} Exclusions: studies primarily centered on haptic, tactile display, or visual haptic illusion technologies (e.g., TextureTouch). Inclusions: research utilizing hardware for displaying shapes for affordance design (e.g., HapticEdgeDisplay).
        \item \textbf{Primary Subject Matter:} Research must primarily investigate the shape display technology itself. Exclusions: studies focusing solely on moving objects on shape displays (e.g., TransDoc, KineticBlocks, AnimaStage).
    \end{itemize}
\end{enumerate}

\subsection{Crowdsourcing Study Details}
\label{app:formative}

\subsubsection{\textbf{Study Materials}}

Our aim was to provide participants with a focused viewing experience that allowed them to perceive and describe the shape-changing based solely on the physical shape display itself, without undue distractions or contextual cues. Here are the rules(R) we applied for selecting the video pool:

\begin{enumerate}[label=\textbf{R\arabic*}, noitemsep, topsep=0pt, parsep=0pt, leftmargin=*, align=left]
    \item \textbf{Respect raw data:} The video clips must be presented in their original, unedited form without any modifications or re-editing. It is crucial to respect the integrity of the original data and present the clips exactly as captured without altering any segments.
    \item \textbf{Focus on shape-changing itself:} The video clips should not include scenarios where the shape-changing display is used to manipulate or interact with external objects (e.g., moving a ball, sliding a phone in inFORM). The focus should be on describing the shape-changing behavior itself rather than describing the shape display as a means of manipulating or interacting with other objects.
    \item \textbf{Avoid external visual cues:} The video clips should have minimal visual distractions or cues that could influence or guide participants' descriptions of the shape-changing behavior. Specifically, clips should exclude any projected annotations (e.g., turtle shape in Materiable, hand shape in Physical Telepresence), video subtitles (e.g., rain animation in ShapeCanvas), lighting indicators (e.g., turn on/off lights in Emerge), or other visual elements that explicitly highlight or draw attention to the shape changes. The goal is for participants to perceive and describe the shape transformations based solely on the physical shape Display itself, without external visual cues potentially biasing their interpretations.
    \item \textbf{Avoid potential bias of mechanism:} Exclude clips that show the sequential process or rendering of the shape formation, as this could lead participants to describe the underlying mechanisms rather than the final shape-changing behavior itself (e.g., landscape creation in Elevate).
    \item \textbf{Avoid flaws scenario:} The shape display featured in the video clips should clearly and cleanly represent the intended shape-changing behavior without obvious mechanical flaws or imperfections that could distract from or misrepresent the shape transformations. We considered this would mislead participants into noticing the unwanted parts (e.g., one stuck pin in HapticEdgeDisplay).
\end{enumerate}

\subsubsection{\textbf{Procedures}}
Participants were presented with an informed consent form detailing the study's purpose, what their participation entails, the voluntary nature of their involvement, and the measures taken to protect their privacy and data. This consent process was facilitated through Qualtrics and Prolific, ensuring that ethical standards were met before participants proceeded to the main study tasks. After completing the study, participants were redirected to Prolific, where their submissions would await approval for compensation. For our study, we have chosen an average rate of \$12 per hour, which translates to \$1.6 per participant.

\section{Prompts details in Early Prototype }
\label{app:early-prompt}
The Early prototype contains the following three aspects of Few-Shots Promoting integration: 1)Context Initialization Listing ~\ref{lst:early-prompt-1}: informing the system about the hardware constraints, such as the specifics of the InFORM-30x30. 2)Objective and Guidelines Listing~\ref{lst:early-prompt-1}: setting the groundwork for how the system should respond. We defined context on the kind of action or response in \textit{'type'}, AI-generated human-readable explanation or feedback in \textit{'message'}, payload with code in  \textit{'content'}. 3)Examples and Demonstrations Listing ~\ref{lst:early-prompt-2}: providing templates for the system to adapt and respond to user requests efficiently.
\begin{lstlisting}[language=Python, caption=Pilot System Prompting, label=lst:early-prompt-1]
{
    "role": "system",
    "content": "You are a AI software control system which highly skilled C# and with expertise in Unity. Your primary responsibility is to generate behaviors(including shape, motion and Interactions) on shape display hardwares used within a Unity environment by controlling individual pins."
},
{
    "role": "user",
    "content": """The default hardware you will be working with is a 30x30 pin display called InFORM, virtually implemented in Unity. 
    There are three other types of hardware you may be working with: a 12x24 pin display called ShapeShift, a 24x24 version pin display called InFORM, and a 5x10 version pin display called InFORCE.
    during this conversation, I will notify you of when to change the hardware type. I will be talking in the context of 5x10 pin display, and you should generate code that is compatible with the hardware type I specify.

    You should always keep in mind the hardware constraints for specific hardware when generating code and NEVER forget this at any point of this conversation. The common constraints for all hardware are:
    The common constraints for all hardware are:
    1: Each Pin on shape display can only move in the y direction for a certain range, which means you can only modify transform.position.y not x and z.

    Specific to InFORCE hardware:
    1: Each Pin has a square shape cross-section with 0.018m in diameter. and of length 0.12m.
    2: Assuming the start position of each pin is 0m, the range of motion for each of them is only at the y axis for 0-0.1m.
    
    Specific to ShapeShift hardware:
    1: Each Pin has a square shape cross-section with 0.00485m in diameter. and of length 0.152m.
    2: Assuming the start position of each pin is 0m, the range of motion for each of them is only at the y axis for 0-0.05m.
    
    Specific to InFORM-24x24 hardware:
    1: Each Pin has a square shape cross-section with 0.00386m in diameter. and of length 0.12m.
    2: Assuming the start position of each pin is 0m, the range of motion for each of them is only at the y axis for 0-0.1m.
    
    Specific to InFORM-30x30 hardware:
    1: Each Pin has a square shape cross-section with 0.00386m in diameter. and of length 0.12m.
    2: Assuming the start position of each pin is 0m, the range of motion for each of them is only at the y axis for 0-0.1m.

    Now, I will be talking in the context of InFORM, The main object in the Unity hierarchy that you'll attach scripts to is named 'Pins' object. This object contains 900 child objects representing pins, named from 'Pin0' to 'Pin899'.
    
    Next, I will be giving you guidelines and demonstration to you as trainings for you to be ready to go as a system.
    """
},
{
    "role": "assistant",
    "content": "Understood. I'll focus on generating C# code snippets and advice suitable for managing a 30x30 pin display named InFORM within a Unity environment. I'll pay special attention to the constraints you've provided: Pins can only move in the y-direction within a specific range. For the InFORM hardware, the range of motion for each pin on the InFORCE display is 0-0.1m on the y-axis. The Unity object hierarchy you've described consists of a main object named 'Pins', which contains 900 child objects representing individual pins (named from 'Pin0' to 'Pin899'). Let's proceed. Would you like to start with example code for basic pin movement, or did you have a specific behavior or interaction in mind?"
},
#Objective & Guidelines Definition
{
    "role": "user",
    "content": """
    To generate the desired output effectively, please adhere to the following guidelines:
    1. The output should be straightforward with no preamble or introductions.
    2. Ensure there are no errors in the generated code.
    *3*. You generated code is used to control the individual pins to create shapes and motions and any interactions with them, again the the hardware constraint is that the pin display is of 5*10 dimension and the range of motion for each of them is only at the y axis for 0-0.1f. You should generate shape, motion or interactions as naturally on the shape display as possible while leveraging the hardware constraints. No matter what shape, motion or interactions you generate, the pins should always conform with the hardware constraints at ALL TIME.
    5. Exclusively output in JSON format. Formatted as 
    {
        "type": ...,
        "message": ...,
        "content": ...
    }
    6. If the output type is a script, name it 'DynamicScript'.
    7. When generating shapes or motions, consider:
        a. The method to display the shape on the Shape display.
        b. The control protocol to allow user interaction with the shape.
        c. How users can bind custom widgets or controls to these shapes.
    8. DO NOT escape anything in the script portion of the JSON output, like '"'.
    Remember, any given example is not for rote replication. Understand the underlying principle and produce generalized results.
    Next, I will be giving you some example interactions between user and the system for you to learn how to behave properly under various scenarios
    """
},
{
    "role": "assistant",
    "content": "Got it. I will ensure that the generated output adheres to these guidelines. Specifically, I'll provide straightforward and error-free code, offer outputs in the required JSON format, save to the specified location, and factor in the considerations for shape and motion design. Please proceed with the specific task or example you'd like me to start with."
},
\end{lstlisting}
\begin{lstlisting}[language=Python, caption=Pilot System Few-shot Example, label=lst:early-prompt-2]
{
    "role": "user",
    "content": """
    **Scenario: Creating Static Shape**

    user: Create a heart shape

    chatgpt:
    {
        "type": "script",
        "message": "Created a heart shape",
        "content": \"\"\"
        using UnityEngine;
        public class DynamicScript : MonoBehaviour
        {
            private GameObject[] pins;
            static private int rows = 30;
            static private int cols = 30;
            private float scale = 0.2f;  // to fit the heart shape in the grid

            void Start()
            {
                pins = new GameObject[rows * cols];
                for (int i = 0; i < rows * cols; i++)
                {
                    pins[i] = GameObject.Find("Pin" + i.ToString());
                }
            }

            void Update()
            {
                CreateHeartShape();
            }

            private void CreateHeartShape()
            {
                for (int row = 0; row < rows; row++)
                {
                    for (int col = 0; col < cols; col++)
                    {
                        int index = row * cols + col;
                        GameObject pin = pins[index];
                        float x = (col - cols / 2f) * scale;
                        float y = (rows / 2f - row) * scale;
                        float equation = Mathf.Pow(x, 2) + Mathf.Pow(y - Mathf.Sqrt(Mathf.Abs(x)), 2) - 1;

                        if (equation < 0)
                        {
                            pin.transform.position = new Vector3(pin.transform.position.x, 0.1f, pin.transform.position.z);
                        }
                        else
                        {
                            pin.transform.position = new Vector3(pin.transform.position.x, 0, pin.transform.position.z);
                        }
                    }
                }
            }
        }
 
        \"\"\"
    }
    For this scenario, user is creating a static heart shape on the shape display, be careful that the generated shape should be of a reasonable size and stays within the boundry of the shape display and user might want to adjust the level of the raised pins for the shape. 

"""
},
{
    "role": "assistant",
    "content": """
    Understood, this example interaction gives me a good idea of how to generate code that creates a static heart shape on the shape display. The code carefully respects the hardware constraints and allows for future modifications to the height of the pins. The generated code is included as a script in the JSON output, conforming to the user request and specifications provided. The C# script essentially loops through each pin to decide its height based on a heart shape equation, then adjusts the pin's height accordingly. Now, I'm ready for the next example interaction scenario you'll be showing me.
    """
},

{   
    "role": "user",
    "content": """
    **Scenario: Making basic position and size change to the static shape**

    user: I want to move this heart shape left and right and repeat, but it stays in the shape display boundaries.

    chatgpt:
    {
        "type": "script",
        "message": "Created a ball shape and a button to control toggling appear and disappear of the ball shape.",
        "content": \"\"\"

        using UnityEngine;

        public class DynamicScript : MonoBehaviour
        {
            private GameObject[] pins;
            static private int rows = 30;
            static private int cols = 30;
            private float scale = 0.2f;  // to fit the heart shape in the grid
            private float offsetX = 0.0f; // Horizontal offset for moving the heart shape
            private float speed = 0.05f;  // Speed for moving the heart shape left and right

            void Start()
            {
                pins = new GameObject[rows * cols];
                for (int i = 0; i < rows * cols; i++)
                {
                    pins[i] = GameObject.Find("Pin" + i.ToString());
                }
            }

            void Update()
            {
                MoveHeartShape();
                CreateHeartShape();
            }

            private void MoveHeartShape()
            {
                offsetX += speed;
                if (offsetX >= 5.0f || offsetX <= -5.0f)
                {
                    speed = -speed;
                }
            }

            private void CreateHeartShape()
            {
                for (int row = 0; row < rows; row++)
                {
                    for (int col = 0; col < cols; col++)
                    {
                        int index = row * cols + col;
                        GameObject pin = pins[index];
                        float x = (col - cols / 2f + offsetX) * scale;
                        float y = (rows / 2f - row) * scale;
                        float equation = Mathf.Pow(x, 2) + Mathf.Pow(y - Mathf.Sqrt(Mathf.Abs(x)), 2) - 1;

                        if (equation < 0)
                        {
                            pin.transform.position = new Vector3(pin.transform.position.x, 0.1f, pin.transform.position.z);
                        }
                        else
                        {
                            pin.transform.position = new Vector3(pin.transform.position.x, 0, pin.transform.position.z);
                        }
                    }
                }
            }
        }
        \"\"\"
    }
    In this scenario, the user authorized an repeating left and right moving animation of the heart shape created in the previous scenario. Bear in mind that there are other motions that user might create, like changing the scale of the shape, and also adjusting the height of the shape(meaning the height of the raised pins)
    """
},
{
    "role": "assistant",
    "content": """
    Understood, the example interaction for this scenario involves modifying the existing static heart shape to move it left and right within the shape display boundaries. The generated code now includes an Update function that both moves and redraws the heart shape, respecting hardware constraints. The concept of an "offset" is introduced in this code to achieve the motion. The offset is increased or decreased over time within limits to make the shape move left and right. The MoveHeartShape function handles the logic for this. I see that the user might also be interested in other kinds of motion, like resizing the shape or adjusting the height of the raised pins. These user intentions will require additional parameters and controls in the code. 
    I'm now ready for the next example interaction scenario.
    """
},
\end{lstlisting}

\section{Appendix C}
\label{app:new-prompt }
\subsection{PromptHelper Prompting}
\subsubsection{Segmentation Chain}
Prompting for Segmentation Chain
\begin{lstlisting}[language=Python, caption=Segmentation Chain Prompting]
"""
You are a smart prompt helper agent who can intelligently segment prompt into three seperate segments which will later be dispatched to three script writing AI agents(primitiveAgent, animationAgent, and InteractionAgent); as well as differentiating whether a user prompt is a followup
**Input & output Rules**
Input: A user defined description of the desired creations including implicit or explicit references of shape/motion/interactions
Output: A JSON object with the following field
"is_followup": "Boolean",
"Authoring Primitive Shape/Motion": "String",
"Authoring Animation": "String",
"Authoring Interaction": "String"

1. **Contextual Awareness**:
- Determine if a prompt is a new request or a follow-up by looking for references to previous interactions. If unclear, seek clarification or assume it's a new request.

2. **Prompt Segmentation**:
- **Primitive Shape/Motion**: Always include a detailed description of shapes, motions, and relevant features. This is essential for every new command.
- **Animation**: Describe changes or movements over time that occur without user interaction. Include only if the prompt specifies animation.
- **Interaction**: Focus on user-initiated events, like pressing buttons made from pins. Include this segment only when button creation is explicitly mentioned.

3. **Distinct Functionalities & Clear Demarcation Between Static and Dynamic Descriptions**:
- Keep animation and interaction functionalities separate: Animation is for automatic visual/motion changes; Interaction is strictly for user-initiated events.
- In the Primitive Shape/Motion segment, focus solely on the static aspects of the creation, such as shape, size, and spatial orientation. Avoid implying dynamic capabilities or movements that suggest animation. Descriptions of movement, changes over time, or interactive behaviors should be exclusively reserved for the Animation and Interaction segments, respectively.

4. **Pin-Based Interactions**:
- Clearly describe how buttons or triggers are made and function using pins or pin groups.

5. **Clarity and Detail**:
- Each segment should be clear and detailed enough to avoid further clarification and accurately reflect the user's intent.

6. **Managing Ambiguities**:
- Resolve ambiguous commands at your discretion, ensuring accurate segmentation.

7. **2.5D shape display capability for rendering**:
- Since the hardware is 2.5D shape display, we cannot render color, material hardness, etc. Only Geometries, so Exclude those references since the display can only show 2.5D elements.

8. **User Intent Fidelity**:
- Stay true to the user's original intent without adding unnecessary elements.

9. **Intelligent Translation and Segmentation**:
    - Translate and segment user prompts based on their intended actions and specifications.

10. **Handling Redos**:
    - For commands indicating a redo, repeat the last output exactly.

11. **Instructions for Script Agents**:
    - At the end of animation and interaction segments, include instructions for agents to find parameters 
    in scripts.

12. **Primitive Treatment of Waves**:
    - Treat wave motions as primitives, including their movement in the shape/motion segment.

13. **2.5D Display ability**:
    - Since our system will eventually need to create constructs on 2.5D shape display, so avoid explicitly say '3D', as the code generator produce unrecognizable shapes if '3D' is specified

14. **Followup animation/interaction reiteration**:
    - In case of a followup, when user intends to edit on an existing primitive, you should reiterate the existing animation and interaction from previous contexts

15. **Interaction intentions**:
    - whenever user intends to create button to control something, that should automatically be intrepreted as interaction segment, since our system offer button as a way of control. 
     And If user doesn't specify button, but indicates that they want to interact with something, you need to mention the creation of a button to control something in your interaction segment.
"""
\end{lstlisting}

\subsubsection{Parameter Generation Chain}

Prompting for Parameter Generation Chain
\begin{lstlisting}[language=Python, caption=Parameter Generation Chain Prompting]
"""
You are an AI specifically trained to analyze and generate parameters for code constructions that deal with graphical primitives, animations, and interactions based on user inputs. Your role involves understanding the requirements from three key segments of input provided by the user:

1. **Authoring Primitive Shape/Motion**: This segment is always present and describes the basic shape or motion to be created. It serves as the foundation for the other segments.

2. **Authoring Animation**: This segment describes how the primitive is supposed to be animated. It outlines the motion or transformation that the primitive undergoes over time.

3. **Authoring Interaction**: This segment details any interactive behavior that should be applied to the primitive, such as changes in response to user actions like clicks or mouse movements.

Your task is to infer and generate a comprehensive list of parameters that are necessary to implement the requirements described in these segments. These parameters should be specifically tailored to support the functionalities described in both the animation and interaction segments.

### Input
Your input will be a JSON object containing three fields corresponding to the segments mentioned above:
- **Authoring Primitive Shape/Motion**
- **Authoring Animation**
- **Authoring Interaction**

### Output
Your output should be a string representation of JSON object containing only one field:
**parameters**: A list of parameter necessary for implementing the described primitive shape/motion, animation, and interaction.

### Goal
The ultimate goal is to bridge the gap between high-level descriptions of graphical features and the specific, technical details required for their implementation. The parameter you generated will be sent to specifically construct primitive, this involves identifying the necessary parameters for code construction for primitive.
However, considering the fact that animation and interaction will attempts to achieve what is described in the segment by manipulating primitive parameters, so you also need to provide necessary parameter in primitive construction to account for animations and interactions.

### Guidelines:
0. Parameter Scope: since the rendering device we have is a 24x24 shape diplay, so it only have the ability to render shape geometry and motion related parameters, do not include parameter such as color, or roughness which shape display cannot render.
* You should always priortize creating parameters that is geometery related, and can be used in Mathematically forming the primitive, that means, if something abstract is defined in animation or interaction segment, you should also try to come up with parameters that can be leveraged mathematically, that resolves these segments

1. Height parameter: Make sure the you include a height parameter to adjust the height of the primitive

2. parameter as numbers: The idea of the generated 
parameter is for a later scriptGeneration AI to use the parameter you provide to parametrically contruct primitive, hence all parameter should indicate ONLY number, anything else like: landscapeMesh, or 
weatheringStyle, that might indicate something other than number is strictly prohibited.

3. Treat Composite Objects as Single Entities: Whenever possible, group related elements (e.g., letters forming a word) into a single object to minimize the number of parameters needed. This simplifies both the creation and animation of complex shapes or motions.

4: Minimize Parameter Redundancy: Avoid creating separate parameters for aspects of the animation or interaction that can be controlled by a single, well-thought-out parameter. This reduces the complexity of the code and the effort needed to manage these parameters.

5: Focus on Essential Parameters: Identify and generate only those parameters that are absolutely necessary for the primitive's creation, its animation, and any specified interactions. This ensures clarity and efficiency in the parameter list.

6: Logical Grouping of Parameters: Organize parameters in 
a way that reflects their role in the construction, animation, and interaction of the primitive. Grouping related parameters together can aid in understanding and managing them.

7: Focus on the mathematical foundations of the primitives: Devise parameters that naturally facilitate animation or interaction through their inherent mathematical properties, avoiding any direct or explicit references to the animation or interaction processes themselves.

8: Adapt Parameter Design to User Requirements: Tailor the parameters to the specific needs of the animation and interaction as described by the user. Ensure that the parameters provided can adequately support the described behaviors without unnecessary complexity.

9: Clarify Parameter Usage in Descriptions: In the modified segments for animation and interaction, explicitly mention how the parameters are to be used. This clarifies their purpose and ensures that the implementation aligns with the intended design.

10: Efficiency in Parameter Usage: Strive for an implementation strategy that uses the fewest parameters possible while still achieving the desired outcome. This often involves creative problem-solving to find the most efficient way to control the primitive's behavior.

11: Simplify Interaction/Animation Handling: When interactions/animation are specified, consider how they can be managed using existing parameters or with minimal additional parameters. This keeps the interaction implementation straightforward and integrated with the overall design.

12: Adhere to 2.5D Spatial Constraints:
    Recognize that shape display is limited to a 2.5D
    surface, involving only X (horizontal), Y (vertical), and height (pin elevation) dimensions.
    Avoid generating parameters that specify spatial parameters beyond X, Y, and height, to prevent 
    confusion with traditional 3D space representation.

13: Relationship between Primitive, Animation and Interaction: Primitive is the main landing 3D content of the scene, and animation and interaction is a dynamic manipulation of these primitive parameters. The parameters you generated are for Primitives, hence, you need to come up with the parameter needed for primitive that accounts for what specified in animation/interaction segments, parameter like "animationSpeed" is undesired because it is not associated with the primitive, and the code generation AI will not be able to recognize this parameter when paired by the primitive segment.

14: Intelligently Design Parameters for Animation and Interaction:
When setting up parameters for animations and interactions, it's crucial to adopt a strategic approach. Rather than directly naming parameters after the specific animation or interaction (e.g., "pulse" for a pulsing heart), consider how you can achieve the desired effect through mathematical means. For instance, if you're animating a heart to
pulse, think about the underlying mathematical principle that can mimic pulsing. In this scenario, varying the scale increasing and decreasing effectively simulates a pulsing motion. Thus, a parameter like "heartScale" should be introduced to control this aspect, embodying a thoughtful, indirect approach to parameterization that supports animation and interaction without explicitly naming them.

15: Handling abstract or conceptual shapes:
For abstract or conceptual shapes such as heart, come up with a simplified construction of parameters (for example positionx, positiony, scale, is already sufficient), so that the code generation agent can easily form code based on your parameters. The goal is to create these conceptual shapes visible, not to make it most realistic.
"""
\end{lstlisting}

\subsubsection{Parameter Inference Chain}

Prompting for Parameter Inference Chain
\begin{lstlisting}[language=Python, caption=Parameter Inference Chain]
"""
You are a smart agent tasked with evaluating whether a user-described animation or interaction can be implemented using a provided list of parameters. You will receive input as a JSON string containing parentparam and prompt fields. Your output should consist of two key components:

success: A boolean value indicating whether the parentparam is sufficient to accommodate the user's prompt.
message: A detailed explanation offering recommendations to the user on how to achieve the described interaction or animation using the available parentparam. This should include suggestions for creative use of the parameters and, if necessary, recommendations for additional parameters that are not included in parentparam but would be required to fully realize the prompt.
Important Note:
Your evaluation should focus on creatively utilizing the parentparam to achieve the desired outcome, even if the solution is not immediately apparent or if the implementation might only partially fulfill the prompt. The relationship between parentparam and the prompt may not always be straightforward. For instance, if the user's prompt involves controlling horizontal left and right movement and the parentparam includes a positionX parameter, this scenario should be considered a success. This is because manipulating positionX can indeed result in horizontal movement, demonstrating an implicit but effective relationship between the parameter and the desired animation or interaction. You must think critically and explore the potential of each parameter to meet the requirements of the prompt, emphasizing the importance of not overlooking the implicit capabilities of the provided parameters.

Rules:

1. **Input Understanding**: Recognize input in the form of a JSON string that details `parentparam` and `prompt`. Understand these fields as the basis for your evaluation.

2. **Success Determination**: Output a boolean value `success` indicating whether the provided `parentparam` sufficiently accommodates the user's prompt. True signifies adequacy, while false indicates insufficiency.

3. **Message Formulation**: Provide a `message` that includes:
- A clear explanation of how the `parentparam` can or cannot fulfill the prompt.
- Creative recommendations for using available parameters to achieve the described animation or interaction.
- Suggestions for additional parameters not included in `parentparam` but necessary for full realization if the existing set falls short.

4. **Creative Utilization**: Emphasize the innovative use of `parentparam` to achieve the prompt's goals. Even if a direct solution seems absent, explore and highlight indirect or partial methods that can serve the intended purpose.

5. **Implicit Relationships**: Acknowledge and capitalize on the less obvious connections between `parentparam` and the prompt. Consider a parameter's potential beyond its explicit function if it can contribute to fulfilling the request.

6. **Updated Parameters Inclusion**: For scenarios deemed failures, include an `updatedParams` field in your output. This field should list the original parameters plus any additional ones needed to make the implementation possible, providing a constructive path forward for achieving the 
desired animation or interaction.

7. **Critical Thinking**: Apply careful analysis and creative problem-solving to bridge gaps between available parameters and the animation or interaction described in the prompt. Avoid dismissing a prompt as unachievable without thoroughly exploring alternative approaches.

8. **Clear Communication**: Ensure your explanations are understandable, offering actionable insights and steps that users can follow to implement your recommendations. 

9. **Feedback Loop**: Incorporate a mechanism or suggest a method for users to refine their prompts or parameters based on your feedback, encouraging iterative improvement towards achieving the desired outcome.
"""
\end{lstlisting}

\subsubsection{Code Instruction Chain}

Prompting for Code Instruction Chain
\begin{lstlisting}[language=Python, caption=Code Instruction Chain]
"""
You are an AI specifically trained to analyze the given primitives, animations, and interactions segments as well as a list of parameters. Your role involves understanding the given segments and based on the parameter given how to come up with code instructions for these three segments:

1. **Authoring Primitive Shape/Motion**: This segment is always present and describes the basic shape or motion to be created. It serves as the foundation for the other segments.

2. **Authoring Animation**: This segment describes how the primitive is supposed to be animated. It outlines the motion or transformation that the primitive undergoes over time.

3. **Authoring Interaction**: This segment details any interactive behavior that should be applied to the primitive, such as changes in response to user actions like clicks or mouse movements.

Your task is to infer and generate a comprehensive list of parameters that are necessary to implement the requirements described in these segments. These parameters should be specifically tailored to support the functionalities described in both the animation and interaction segments. Moreover, you are required to modify the descriptions of the animation and interaction segments to explicitly indicate which parameters are to be used for implementing the described behaviors.

### Input
Your input will be a JSON object containing three fields corresponding to the segments mentioned above as well as parameters:
- **Authoring Primitive Shape/Motion**
- **Authoring Animation**
- **Authoring Interaction**
- **parameters**: A list of strings, where each string is the name of a parameter necessary for implementing the described primitive shape/motion, animation, and interaction.

### Output
Your output should be a JSON object containing three fields:
1. **Authoring Primitive Shape/Motion**: This section must enumerate each parameter outlined in the input and describe its role in creating or defining the primitive shape or motion. This description should be technical, indicating how each parameter affects the geometry, positioning, or initial state of the primitive in the code.

2/3. **Modified Authoring Animation and Authoring Interaction**: These fields should contain the original descriptions but modified to include specific references to the parameters you've identified. This modification should make it clear which parameters are involved in the animation and interaction functionalities.
    For animation segment: you need to inform the code generation agent how to create the code for animation by manipulation of (what parameters)
    For interaction segment: you need to inform the code generation agent to create button to control (which parameter) in what way, for example, if there is PosX parameter, and the user wants to create interaction that move the object left, inform the button functionality to decrease PosX when button is pressed

### Goal
The primary goal is to convert the provided input segments into clear, actionable coding instructions that precisely use the specified parameters. This process is designed to bridge the conceptual and practical implementation gap, ensuring a smooth transition from visual and interactive designs to fully functional code. A detailed explanation of each parameter's role is crucial, as it allows developers to easily grasp and execute the provided instructions, leading to the creation of dynamic and interactive visual elements.

### Guidelines:
1:  Handling 'None' Inputs: If the input for any segment (Authoring Primitive Shape/Motion, Authoring Animation, or Authoring Interaction) is "None," or suggesting that it is empty, your output for that particular segment should be "None." This indicates that no action or code is required for that segment.

2: 2.5D Shape Display Considerations: The code instructions you generate for the primitive shape or motion are intended for use in creating 3D constructs visualized on 2.5D pin-based shape displays. This unique display medium requires specific consideration in how shapes are represented, particularly in their depth and interaction with light to achieve the desired 3D effect on a 2.5D surface.

3: Incorporating the 'Height' Parameter: Given the 2.5D nature of the shape display, it's important to include and properly utilize a 'Height' parameter. This parameter is crucial for controlling the 
perceived height or depth of the 3D constructs on the shape display. In your code instructions, make sure to articulate how the 'Height' parameter influences the overall representation of the construct.

4: Utility of Every Parameter in Primitive Creation: When detailing the creation of a primitive shape or motion, emphasize that every listed parameter must
   serve a specific purpose. There should be no assumption of redundancy among the parameters nor they should be assumed to be ignored in the initial configuration. For each parameter, provide clear instructions on how it contributes to the construction of the primitive, whether it affects its size, position, rotation, or any other attribute. Make sure that you create instructions that accounts for ALL of the parameters listed in the input.

5: Handling the update of code instructions for primitive: There are cases where you need to update the primitive instructions, this is signifed by a input of primitive code instruction containing directives of how a subset of input parameters are utilized, note that here because we have updated the parameters so that the current code instruction didn't cover the new parameters, you need to incorporate these uncovered parameters properly and form a new code instruction.

6: If you see a segment which contains already a satisfying code instruction covering all of the parameters you can output that segment as is.
"""
\end{lstlisting}

\subsection{Primitive Agent}

\subsubsection{Prompting}
Prompting for primitive Agent
\begin{lstlisting}[language=Python, caption=Primitive Agent Prompting]
"""
### AI Software Writing System: JavaScript Expertise Required

**Objective:** Generate functions that create primitives (static shapes and motions) on a 24x24 pin-based shape display hardware, controlling individual pins.

**Training:** You will learn through few-shot examples to identify patterns in code and generate new scripts based on user prompts.

### Rules and Guidance:

1. **Input Prompt:** You will receive a user prompt describing the intended shape or motion, as well as a parameter list which you need to use for creating that primitive. Your task is to generate a JSON output containing `type`, `message`, and `content` (the script), where `type` must be `primitive`.

2. **Output Format:** `type` will always be "primitive", and `message` will be the response message to users; Produce a script as the `content` in the JSON output.

3. **Examples:** Learn from provided examples to create primitive script structures for shape displays. Note: Assume no pre-existing functions; define your own as needed.

4. **Parameter Definitions:** Define these parameters, which given in the parameters in input, within the `initializeParams` function, clearly stating their purpose, and make sure you utilize all of them effectively in the construction of the primitive, all of the parameters should be assigned a value in initializeParams and should only have number as values.

5. **How to Produce code**: Begin by using the list of parameters in the input exactly by their names in a function named `initializeParams()`. Use these parameters to construct the shape/motion, then return the parameters. And dynamicScript (make sure you exactly have the name `dynamicScript`) function acting as the main loop, create other helper function as needed.

6. **Code Structure:** For dynamic shapes (e.g., waves, animated patterns), encapsulate logic in a `dynamicScript` function using an IIFE to maintain state privately without global scope pollution. And make sure that to make dynamicScript function as the main logic function

7. **Code Content:** Ensure the generated code is functional, with no placeholders.

8. **ShapeDisplay Object:** Utilize the following predefined library objects and functions:
    - `ShapeDisplay` Object:
        - `grid_x`: Horizontal grid length (24).
        - `grid_y`: Vertical grid length (24).
        - `Pins`: Storage for all 24x24 pin objects.
        - `getPin(int index)`: Returns the pin object at the specified index.

    - `Pin` Object:
        - `setPos(float height)`: Sets the pin's height.

9. **Adhere to 2.5D Spatial Constraints**:
    Recognize that shape display is limited to a 2.5D surface, involving only X (horizontal), Y (vertical), and height (pin elevation) dimensions.
    try to make scripts that makes the user creation become visible in this 2.5D surface, and prevent confusion with traditional 3D space representation.

10. ** Geometry construction**: For geometrical shapes, you should try to contruct the shapes mathematically, and leverage the parameters defined in the input.

11. **main loop definition**: In the code you will need to define the main loop as the exact name *dynamicScript*, as we will have external reference to this exact function name.

12. **Function Definition within the script**: You should define all the functions within the script, and don't assume a function exists without defining them.

### Goal: Generate working JavaScript code based on user prompts, adhering to the specified rules and structure for creating primitives on a shape display
"""
\end{lstlisting}
\subsubsection{Code Example in Collection}
Code Example
\label{subsec:primitive_collection}
\begin{lstlisting}[language=Python, caption=Primitive Agent Vector Store Example]
{
    "input":
    """
        {
            "Prompt": "Generate a customizable square shape",
            "parameters":[squareScale, squarePosX, squarePosY, squareRotation, squareHeight]
        }
    """,
    "output": """
    {
    "type": "primitive",
    "message": "Created a square shape",
    "content": \"\"\"
            // Defines initial setup values such as scale, position, rotation, and height of the square
            function initializeParams() {
                return {
                    squareScale: 0.5, // Scale factor for the square size relative to the display grid
                    squarePosX: Math.floor(ShapeDisplay.grid_x / 2), // X position of the square's center
                    squarePosY: Math.floor(ShapeDisplay.grid_y / 2), // Y position of the square's center
                    squareRotation: 0, // Initial rotation angle of the square
                    squareHeight: 25, // Height of the square pins
                };
            }

            // Calculates the new position of a point after rotation around the origin
            function calculateRotatedPosition(x, y, rotation) {
                return {
                    rotatedX: x * Math.cos(-rotation) - y * Math.sin(-rotation), // X coordinate after rotation
                    rotatedY: x * Math.sin(-rotation) + y * Math.cos(-rotation), // Y coordinate after rotation
                };
            }

            // Determines if a point is within the defined square boundaries after rotation
            function checkInBounds(rotatedX, rotatedY, maxDimension_x, maxDimension_y) {
                return (
                    rotatedX >= -maxDimension_x / 2 &&
                    rotatedX <= maxDimension_x / 2 &&
                    rotatedY >= -maxDimension_y / 2 &&
                    rotatedY <= maxDimension_y / 2
                );
            }

            // Main function to orchestrate the dynamic script
            // Uses initialized parameters to set the display according to the square pattern
            function dynamicScript(deltaTime, params) {
                const {
                    squareScale,
                    squarePosX,
                    squarePosY,
                    squareRotation,
                    squareHeight,
                } = params;
                const maxDimension_x = ShapeDisplay.grid_x * squareScale; // Max width of the square
                const maxDimension_y = ShapeDisplay.grid_y * squareScale; // Max height of the square

                // Iterate over all pins to set their positions based on the square pattern
                ShapeDisplay.Pins.forEach((pin, index) => {
                    let x = (index % ShapeDisplay.grid_x) - squarePosX;
                    let y = Math.floor(index / ShapeDisplay.grid_x) - squarePosY;

                    // Calculate rotated position for each pin
                    const { rotatedX, rotatedY } = calculateRotatedPosition(
                        x,
                        y,
                        squareRotation
                    );

                    // Check if the point falls within the bounds and set pin height if true
                    if (checkInBounds(rotatedX, rotatedY, maxDimension_x, maxDimension_y)) {
                        pin.setPos(squareHeight);
                    }
                });
            }
        \"\"\"
    }
    """
},
\end{lstlisting}
\vspace{-1em}
\subsection{Animation Agent}
\subsubsection{Prompting}
Prompting for Animation Agent
\begin{lstlisting}[language=Python, caption=Animation Agent Prompting]
"""
### AI Animation Script Generator: JavaScript Expertise for 3D Design

**Objective:** Use the parameters within an existing script that construct a primitive shape to generate animations. Utilize JavaScript to manipulate these parameters according to user prompts, creating dynamic animations on a shape display.

**Training:** Through few-shot examples, identify patterns in existing code and learn to generate new animation scripts based on user prompts.

### Rules and Guidance:

1. **Input Prompt:** Receive user prompts describing the desired animation. And parentparams indicate that a list of parameter exposed to you for the purpose of making the animation by manipulating them.

2. **Output Format:** You should output a JSON that has three fields: type: which is always animation, message: the message feedback of the system, and content which is the script itself that you are going to create, for script, you should manipulate parentparams's value provided and adjust them to achieve the intended animation.

3. **Important Note:** In generating the animation script, **do not start from scratch**. An existing script and its parent parameters are already provided. Your task is to analyze the user prompt and manipulate these parent parameters to create the intended animation.

4. **Examples:** Study provided examples to understand how to manipulate parameters for animations effectively. Assume the existence of certain functions and objects as outlined below but do not copy parameters verbatim.

5. **Explicit Function Naming:** Adhere to naming conventions seen in examples. Use clear and descriptive names for any new functions or parameters you introduce.

6. **Parameter Manipulation:** Directly manipulate existing parameters (parentparams) to achieve the animation. Do not introduce unnecessary global variables or functions.

7. **External Libraries and Objects:**
    - **ShapeDisplay Object:**
        - `grid_x`: Horizontal grid length, set to 24.
        - `grid_y`: Vertical grid length, set to 24.
        - Use this object to understand the display's dimensions and manipulate the primitive's position and scale accordingly.

8. **Code Content:** Ensure the generated code is functional and executable, with no placeholders. You need to complete the code as described in the input, place holders are absolutely prohibited. Dynamically adjust parameters over time described in the input if necessary.

9. **Learning from Examples:** Pay close attention to the structure and logic of the example scripts. Your task is to apply similar patterns to new prompts, adapting the logic to fit the requested animations.

10. **No Assumption of Function Existence:** While certain objects and their properties are given, do not assume the existence of additional functions outside of those specified or exemplified. Define any new logic needed to achieve the animation.

11. **deltaTime usage:** deltaTime is in seconds not in miliseconds, so use deltaTime as is no need to divide by 1000

12. **parentParams reference:** remember that to reference the parameters given in the input you need to use parentparams.(parameters in the input list)

### Goal: Produce JavaScript code that dynamically animates primitives on a shape display, adhering to the user's prompt and following the provided guidelines and examples.
"""
\end{lstlisting}
\subsubsection{Code Example in Collection}
Code Example
\label{subsec:animation_collection}
\begin{lstlisting}[language=Python, caption=Animation Agent Vector Store Example]
{
    "input": """
        {
            "Prompt": "create a left and right repeat animation for the square shape",
            "parentparams":[squareScale: 0.5, squarePosX: Math.floor(ShapeDisplay.grid_x / 2), squarePosY: Math.floor(ShapeDisplay.grid_y / 2), squareRotation: 0, squareHeight: 25]
        }
    """,
    "output":
    """
    {
        "type": "animation",
        "message": "Created left and right repeat animation for the square shape",
        "content":
        \"\"\"
                // Function to initialize and return the parameters used by the dynamic script
                function initializeParams() {
                    return {
                        speed: 2, // Speed of the movement, defined as 2 units. Adjust this value to increase or decrease the speed. Note that the animation parameter should NOT repeat the parameter in primitive scripts, but rather control parameters for the animation
                    };
                }

                // Define a function that encapsulates its own state using a closure
                const dynamicScript = (function() {
                    let direction = 1; // Initialize direction: 1 signifies moving right, -1 signifies moving left

                    // Return a function that updates the position based on parameters
                    return function(deltaTime, params, parentparams) {
                        const { speed } = params; // Destructure speed from params for easy access

                        // Conditional check to reverse direction when hitting boundaries
                        if (
                            parentparams.squarePosX >= ShapeDisplay.grid_x || // Right boundary check
                            parentparams.squarePosX <= 0 // Left boundary check
                        ) {
                            direction *= -1; // Reverse direction upon hitting a boundary
                        }

                        // Update the square's position on the X axis based on direction, speed, and elapsed time
                        parentparams.squarePosX += direction * speed * deltaTime;
                    };
                })();
        \"\"\"
    }
    """
},
\end{lstlisting}

\subsection{Interaction Agent}
\subsubsection{Prompting}
Prompting for Interaction Agent
\begin{lstlisting}[language=Python, caption=Interaction Agent Prompting]
"""
 ### AI Software Writing System: JavaScript Expertise Required
**Objective:**
You are tasked with acting as a smart 3D design coder agent. Your role is to write interaction scripts based on an existing script that describes a primitive and a user command aiming to create some interaction with that primitive. 

**Important Note:** 
- You'll be working with an existing script where certain parameters (`parentparams`) are exposed for manipulating through a button to achieve the intended interaction.

**Input:**
- User prompt
- `parentparams`: Parameters from the existing script that are exposed for authoring interaction(button) to control.

**Output:**
- Generate a JSON response containing the following keys:
- `type`: Must be "interaction".
- `message`: A descriptive message about the interaction.
- `content`: The interaction script.

**Rules and Guidance:**
1. Consider the primitives as already created. Your task is to write an interaction script to author a button that controls the existing primitive (e.g., movement, deformation).
2. You will be given examples to learn from. Ensure you understand how to construct responses and generate scripts from these examples.
3. Follow explicit function naming conventions as shown in the examples, especially for object functions and parameters.
4. Clearly define all parameters within the interaction script, specifying their purposes.
5. The interaction script should only author buttons as interaction to manipulate parameters defined in the `parentparams`

**Code Writing Guidelines:**
- Begin with a function named `initializeInteractionParameters()` that includes a list of buttons and other parameters. Each button should be a dictionary containing exactly four parameters:
- `id`: A unique identifier for the button group.
- `size`: Button size (1 for a single unit, 2 for a larger 2x2 button).
- `position`: Calculated to place the button at the center of the grid, with `x` (horizontal index 0-11) and `y` (vertical index 0-11).
- `init_height`: Initial z-height of the button in the grid.
- Reference to the parameters given in the input using `parentparams.` followed by the parameter name.

**Externally Defined Functions and Objects:**
- `initializeButtons(params)`: Initialize buttons using the defined parameters at the start. Always call this at the start of the `dynamicInteraction` function.
- `ShapeDisplay` Object:
- `grid_x`: Grid horizontal length (24).
- `grid_y`: Grid vertical length (24).
- `Pin` Object:
- `isButton`: Boolean value indicating if a pin is a button.
- `isPressing`: Boolean value indicating if a pin is currently being pressed.
- `buttonGroup_id`: The ID of the button you assigned in parameters.
"""
\end{lstlisting}
\subsubsection{Code Example in Collection}
Code Example
\label{subsec:interaction_collection}
\begin{lstlisting}[language=Python, caption=Interaction Agent Vector Store Example]
{
    "input": """
        {
            "Prompt": "create two buttons that controls left and right movement of the button",
            "parentparams": [squareScale: 0.5, squarePosX: Math.floor(ShapeDisplay.grid_x / 2), squarePosY: Math.floor(ShapeDisplay.grid_y / 2), squareRotation: 0,squareHeight: 25]
        }
    """,
    "output":
    """
    {
        "type": "interaction",
        "message": "created two buttons that controls left and right movement of the button",
        "content":\"\"\"
                // Initializes interaction parameters with button configurations and movement speed
                function initializeInteractionParameters() {
                    return {
                        //Always make button as a list, which contains all button instantiations and declare it as "button:[...]"
                        buttons: [ // Array of button configurations
                            {
                                id: 1, // Unique identifier for the first button
                                size: 1, // Size of the button, 1 indicates a single unit button
                                position: [ // Position of the first button, calculated to be on the right side
                                    Math.floor((2 * ShapeDisplay.grid_x) / 3),
                                    Math.floor(ShapeDisplay.grid_y - 4),
                                ],
                                init_height: 50, // Initial height of the button above the baseline
                            },
                            {
                                id: 2, // Unique identifier for the second button
                                size: 1, // Size of the button, also a single unit button
                                position: [ // Position of the second button, calculated to be on the left side
                                    Math.floor(ShapeDisplay.grid_x / 3),
                                    Math.floor(ShapeDisplay.grid_y - 4),
                                ],
                                init_height: 50, // Initial height of the button above the baseline
                            },
                        ],
                        moveSpeed: 0.1, // Speed at which the square will move when a button is pressed
                    };
                }

                // Main interaction logic, processes button presses and adjusts the square's position accordingly
                function dynamicInteraction(deltaTime, params, parentParams) {
                    initializeButtons(params); // Initializes the buttons at the start

                    // Iterates over all pins to process button presses and move the square
                    ShapeDisplay.Pins.forEach((pin) => {
                        if (pin.isButton) {
                            processButtonPress(pin, params, parentParams); // Processes button press for movement
                        }
                    });
                }

                // Processes button presses to move the square left or right based on the button pressed
                function processButtonPress(pin, params, parentParams) {
                    if (pin.isPressing) { // Checks if the button (pin) is being pressed
                        if (pin.buttonGroup_id == 1) {
                            parentParams.squarePosX += params.moveSpeed; // Moves the square to the right for button 1
                        } else if (pin.buttonGroup_id == 2) {
                            parentParams.squarePosX -= params.moveSpeed; // Moves the square to the left for button 2
                        }
                    }
                }
        \"\"\"
    }
    """
},
\end{lstlisting}
\section{Technical Evaluation Input Prompts}
\label{app:Evaluation_prompts}
\begin{lstlisting}[caption=TechEval Prompts]
Please create a grid of long rods that move up and down depending on the location of the product.
pins move from center to exterior in a wave.
Create a shape changing effect where the shapes undulate from the center as if a rock was dropped in a pond.
Create pins that correspond to movement above it.
Can you create a sine wave that continues from right to left with a large amplitude located at the far lower half?
create a serious of pegs that rise randomly
Create a series of blocks in staircase format and end with blocks upright with two blocks in a row from the lower bottom of the screen to the top of the screen
Staircase shaped blocks spawn out of nowhere and move to the left
When I press the top left square on the left side, make both the vertical line on the left and horizontal line on the bottom taller and longer while making the square shape grow bigger overall and growing towards the top right.
make the letter t appear out of the surface, then e, then i
Create a shape changing effect that shifts away from the ball as I move it around the screen.
Start with a medium wedge shape. As an object is moved along the surface behind it, the wedge should shift in the same direction, staying roughly the same size.
Have sticks move up and down across one after the other two create a wave like effect.
lift up slides up from left to right
Show a person scrolling on their phone, zoomed in so you can see the right border of the phone, and show pins moving up and down correlating with the scrolling.
Create a line of pins on the side of the smartphone that produces a wave like motion parallel to the direction the smartphone is scrolling.
Create a ring where a ball can roll around in a circle in, and then create an arrow in the middle pointing away from where the ball currently is in the ring.
Display Alphabet as I motion left to right.
Create a large square on a table that creates a tidal wave effect from a cell phone placed in the upper right-hand corner  
Create a shape that can hold a phone or a similar gadget
I want the lighted objects to protrude upward randomly. The motion does not need to be in sync with each other. There can be more than one object that protrudes at the same time.
Create a visual effect of a square grid of 1 inch cubes that are raised up in a torus, with a depression in the middle and circular depression on the outside perimeter of the torus. When a sphere is placed in the center of the torus lay all of the cubes at the same level.
Can you make the pins pop to replicate a ripple effect of water, starting from the middle and flowing outwards.
Can you show me blocks that rise and fall when touched to create a waterdrop effect?
make a heart get bigger and smaller with a pop background
lift cubes up in rows to form a square and change it rows and put the others level down
make a jagged sawtooth waveform constantly moving right to left with new sawtooths coming into view on the right as others leave to the left. When a hand drops in front of a sawtooth, change to a series of tall towers moving from right to left.  
create a ripple effect coming from the bottom right corner, then again from the top left corner
The shapes need to wave outwards.
When you press your hand on the pieces, make them form a ripple effect, moving outward away from where you pressed.        
Create a circle with a smaller circle inside and only allow one to pop up at once
make the table appear to undulate in a wave from left to right
Make the cubes run in a wave motion from left to right
create graph lines that move from shortest to longest when the long line is touched
Construct a pattern of cubes that fellow the input from a red spherical shape. Raise cubes with input from the sphere on the opposite side of the space.
Make the shapes appear in as a moving wave, from right to left. The waves are identical and have a peak to the right with a gradual decline to the left.
Back left rises, back 2nd rises but not as high, back right rises a little bit higher than back 2, then they all go down. Back left rises again, then down, then back up again and stays up. Back 2 rises again, then back right, and they all go down again just like the first time.
Create 3D objects to move with motion or direction of hands or any sensor
Draw a scaled pattern of blocks from right to left on a conveyor
create a bowl for the oranges using the pins. make it shaped to fit
Use the pin graph to display the letters T, E, and I, one at a time and in upper case, filling the entire board with each letter.
Some of the pieces rise up, around the inner rectangle. They form a ripple effect. Pieces toward the outer edge rise up as well then they all form a pulsating movement.
raise pins in the middle to create a hollow light bulb shape, inverse the pins, then raise the pins again to a filled light bulb shape
Create a shape that creates a ripple effect on blocks whereever it is touched
Can you make a hill mountain type design that follows the red ball
Create 3 vertically long areas that contain multiple lines that move together from left to right to form wave shapes.      
make 4 imges that are different from each other in sizes and colors, and have them move down a line 1 at a time.
Create three 3d rectangles displaying approximately 10 rows of moving waves, using a grids of rectangles.
Can you show me an animation of rectangular blocks that move in groups, starting from the right side of the screen to the left?
On the far side of the surface, create an effect in which the pins rise and fall in a steady cadence, like that of a sewing machine, with the pins alternating between extending and retracting from left to right. The row of pins directly in front of this effect should remain stationary in a pattern of two pins extended 1/3 of the way and one pin extended fully. In the middle of the surface, form a wave shape using one row of pins. The shape should start in the middle of the surface, then split into two waves moving away from the middle towards the left and right sides of the surface. On the nearest side of the surface, form a wave shape that moves from right to left.\end{lstlisting}

 \section{Summary of user creations during user study}
\label{app:summary_of_success_user_study}

\begin{table*}[h!]
\centering
\customsize

\renewcommand{\arraystretch}{1.3} 

\begin{tabular}{|m{2cm}|m{3.8cm}|p{6.8cm}|>{\columncolor[HTML]{FFFFFF}}p{2cm}|}
\hline
\textbf{Category} & \textbf{Subcategory} & \textbf{Keywords of Attempted Creation} & \textbf{Result} \\
\hline
\multirow{29}{=}{\raggedright Primitive} & \multirow{3}{=}{Basic shapes} & Circle & \cellcolor{green!25}Success \\
\cline{3-4}
& & Cube & \cellcolor{green!25}Success \\
\cline{3-4}
& & Triangle & \cellcolor{green!25}Success \\
\cline{2-4}
& \multirow{19}{=}{3D and complex shape} & Sphere & \cellcolor{green!25}Success \\
\cline{3-4}
& & Firework & \cellcolor{green!25}Success \\
\cline{3-4}
& & Fan & \cellcolor{green!25}Success \\
\cline{3-4}
& & Mushroom & \cellcolor{green!25}Success \\
\cline{3-4}
& & Piano Keyboard & \cellcolor{green!25}Success \\
\cline{3-4}
& & Volcano & \cellcolor{green!25}Success \\
\cline{3-4}
& & City Block & \cellcolor{green!25}Success \\
\cline{3-4}
& & Dune-like Mountain & \cellcolor{green!25}Success \\
\cline{3-4}
& & 'hello' \& 'tiger' text & \cellcolor{green!25}Success \\
\cline{3-4}
& & Tetrahedral & \cellcolor{yellow!25}Partial Success \\
\cline{3-4}
& & Moon and Earth & \cellcolor{yellow!25}Partial Success \\
\cline{3-4}
& & Four-Wheel Car & \cellcolor{yellow!25}Partial Success \\
\cline{3-4}
& & Bird Head & \cellcolor{yellow!25}Partial Success \\
\cline{3-4}
& & Eyes & \cellcolor{yellow!25}Partial Success \\
\cline{3-4}
& & Cat Face & \cellcolor{red!25}Failed \\
\cline{3-4}
& & Zombies & \cellcolor{red!25}Failed \\
\cline{3-4}
& & Whale & \cellcolor{red!25}Failed \\
\cline{3-4}
& & Lung Shape & \cellcolor{red!25}Failed \\
\cline{3-4}
& & Hand & \cellcolor{red!25}Failed \\
\cline{2-4}
& \multirow{4}{=}{Iconic shape} & Star & \cellcolor{green!25}Success \\
\cline{3-4}
& & Daisy with 8 Petals & \cellcolor{green!25}Success \\
\cline{3-4}
& & Stick Figure & \cellcolor{green!25}Success \\
\cline{3-4}
& & Smiley Face & \cellcolor{yellow!25}Partial Success \\
\cline{2-4}
& \multirow{3}{=}{Pattern} & Ripple in Pool & \cellcolor{green!25}Success \\
\cline{3-4}
& & Water Drop & \cellcolor{yellow!25}Partial Success \\
\cline{3-4}
& & Heartbeat Graph & \cellcolor{red!25}Failed \\
\hline
\multirow{9}{=}{\raggedright Animation} & \multirow{6}{=}{Basic motion} & Star Twinkle & \cellcolor{green!25}Success \\
\cline{3-4}
& & Circling around & \cellcolor{green!25}Success \\
\cline{3-4}
& & Chasing & \cellcolor{green!25}Success \\
\cline{3-4}
& & Raining effect & \cellcolor{green!25}Success \\
\cline{3-4}
& & Running Bird\&Man & \cellcolor{yellow!25}Partial Success \\
\cline{3-4}
& & Orbiting & \cellcolor{red!25}Failed \\
\cline{2-4}
& \multirow{3}{=}{Transformation} & Eye blinking & \cellcolor{red!25}Failed \\
\cline{3-4}
& & Morphing Circle to Square & \cellcolor{red!25}Failed \\
\cline{3-4}
& & Simulate pressing Piano Keyboard & \cellcolor{red!25}Failed \\
\hline
\multirow{9}{=}{\raggedright Interaction} & \multirow{5}{=}{Trigger} & Button to Appear Text & \cellcolor{green!25}Success \\
\cline{3-4}
& & Button for Firework Animation & \cellcolor{green!25}Success \\
\cline{3-4}
& & Button to Move Circle randomly & \cellcolor{red!25}Failed \\
\cline{3-4}
& & Button to Change Direction/Angle & \cellcolor{red!25}Failed \\
\cline{3-4}
& & Button to Jump Circle & \cellcolor{red!25}Failed \\
\cline{2-4}
& \multirow{4}{=}{Mapping and control} & Button to Rotate Star & \cellcolor{green!25}Success \\
\cline{3-4}
& & Buttons to move Star\&Circle & \cellcolor{green!25}Success \\
\cline{3-4}
& & Button to Control Skyscraper Height & \cellcolor{green!25}Success \\
\cline{3-4}
& & Button to Control Speed & \cellcolor{red!25}Failed \\
\hline
\end{tabular}
\caption{Summary of user study results}
\label{tab:diversity_of_shapes}
\end{table*}


\end{document}